\documentclass[%
 reprint,
 amsmath,amssymb,
 aps,
]{revtex4-2}

\usepackage{graphicx}
\usepackage{subcaption}
\usepackage{dcolumn}
\usepackage{tabularx}
\usepackage{bm}
\usepackage{amsfonts,color} 
\usepackage[table,xcdraw]{xcolor}
\usepackage[english]{babel}

\definecolor{Gray}{gray}{0.9}

\newcolumntype{Y}{>{\centering\arraybackslash}X}
\newcommand*\mean[1]{\bar{#1}}

\begin{document}

\preprint{APS/123-QED}

\title{
Analyzing admissions metrics as predictors of graduate GPA and whether graduate GPA mediates PhD completion
}

\author{Mike Verostek}
\affiliation{
 Department of Physics and Astronomy, University of Rochester 
}
 \affiliation{Department of Physics and Astronomy, Rochester Institute of Technology}
  \email{mveroste@ur.rochester.edu}

\author{Casey Miller}
\affiliation{
 Department of Physics and Astronomy, Rochester Institute of Technology
}

\author{Benjamin Zwickl}
\affiliation{
 Department of Physics and Astronomy, Rochester Institute of Technology
}

\date{\today}

\begin{abstract}
An analysis of 1,955 physics graduate students from 19 PhD programs shows that undergraduate grade point average predicts graduate grades and PhD completion more effectively than GRE scores. Students' undergraduate GPA (UGPA) and GRE Physics (GRE-P) scores are small but statistically significant predictors of graduate course grades, while GRE quantitative and GRE verbal scores are not. We also find that males and females score equally well in their graduate coursework despite a statistically significant 18 percentile point gap in median GRE-P scores between genders.  A counterfactual mediation analysis demonstrates that among admission metrics tested only UGPA is a significant predictor of overall PhD completion, and that UGPA predicts PhD completion indirectly through graduate grades.  Thus UGPA measures traits linked to graduate course grades, which in turn predict graduate completion.  Although GRE-P scores are not significantly associated with PhD completion, our results suggest that any predictive effect they may have are also linked indirectly through graduate GPA.  Overall our results indicate that among commonly used quantitative admissions metrics, UGPA offers the most insight into two important measures of graduate school success, while posing fewer concerns for equitable admissions practices.


\end{abstract}

\maketitle

\section{\label{sec:Introduction}Introduction}
As physics graduate admission committees across the country consider eliminating GRE scores from consideration when evaluating applicants \cite{miller_typical_2019, gre_requirements_googledoc}, it is important to continue examining the GRE's ability to predict success in graduate school in order for programs to make informed policy choices. Although GRE scores are among the numeric metrics that best predict admission into U.S. graduate programs \cite{attiyeh_testing_1997, posselt2019metrics}, there are significant disparities in typical GRE performance between students of different demographic backgrounds \cite{miller_test_2014}.  Combined with the fact that Physics remains one of the least diverse of all the STEM fields \cite{noauthor_nsfgov_nodate}, the prospect that GRE tests limit the ability of certain students to enter graduate school has led researchers to begin questioning the utility of GRE exam scores in the graduate admissions process in comparison to other quantitative metrics such as undergraduate GPA (UGPA) \cite{miller_typical_2019, levesque_physics_2015, young2021physics}.  Among some of the findings in this body of work are indications that earning high marks on the GRE Physics (GRE-P) test fails to help students ``stand out" to admissions committees who would have overlooked them due to an otherwise weak application  \cite{young2021physics}, and that typical physics PhD admissions criteria such as the GRE-P exam fail to predict PhD completion despite limiting access to graduate school for underrepresented groups \cite{miller_typical_2019}.

Yet overall PhD completion is only one measure of ``success" in graduate school.  Graduate faculty often cite high grades, graduation in a reasonable amount of time, and finding a job after graduation as indications of successful graduate students \cite{walpole_selecting_2002}.  It is therefore crucial for admissions committees understand how these other measures of success are related to common quantitative admissions metrics as well.  In particular, studying the role of graduate grade point average (GGPA) is important for both historical and practical reasons.  Among physics graduate students, positive relationships between GRE-P scores, first-year graduate grades, and cumulative graduate grades have traditionally been touted as evidence for the exam's utility in evaluating applicants \cite{gre1989validity, kuncel2001comprehensive}.  Several other studies \cite{petersen2018multi, kuncel2010validity, hall2017predictors, moneta2017limitations} suggest that a number of common admissions metrics are correlated with GGPA as well.  At a practical level, gaining a better understanding of which factors best predict graduate grades is valuable due to the fact that performance in graduate classes can influence whether students will ultimately complete a PhD.  For instance, programs may institute GPA requirements that prevent students from continuing study if their course grades do not meet certain criteria.

Predictive validity analyses of GRE scores across all STEM disciplines consistently find that scores on the GRE Quantitative (GRE-Q) and Verbal (GRE-V) tests are more effective predictors of graduate grades than PhD completion \cite{petersen2018multi, kuncel2010validity}.  For instance, recent studies on PhD admissions in the biomedical field found that students' GRE-Q and GRE-V scores are poor predictors of PhD completion, but are more associated with first-semester and cumulative graduate school grades \cite{hall2017predictors, moneta2017limitations}.  In contrast, studies cited by the Educational Testing Service (ETS) such as the meta-analysis of GRE predictive validity by \citeauthor{kuncel2001comprehensive} \cite{kuncel2001comprehensive} show a positive correlation between GRE subject scores, graduate grades, and PhD completion.  \citeauthor{kuncel2001comprehensive} find that GRE subject tests show larger correlations with GGPA than GRE-Q, GRE-V, or UGPA, which they attribute to the subject-specific knowledge that the GRE subject tests are purported to measure. Still, GRE-Q and GRE-V scores, which the authors presume to be broad measures of cognitive ability, are shown to only moderately correlate with GGPA but do not significantly correlate with PhD completion.  \citeauthor{kuncel2001comprehensive} also find undergraduate grade point average (UGPA) correlates with GGPA but not completion. 



Despite voluminous research on the efficacy of quantitative admissions metrics in predicting graduate success, there remains a dearth of studies specifically examining these metrics in the context of physics graduate education.  No current study elucidates the relationships between undergraduate grades, GRE scores, and physics graduate grades.  Moreover, studies such as \cite{miller_typical_2019} do not incorporate graduate grades into models of PhD completion despite its theoretical and structural importance on the road to graduate success.  This paper aims to fill these gaps in the current literature.

The primary goal of this paper is to extend the analysis of \citeauthor{miller_typical_2019} \cite{miller_typical_2019}, using the same data set to examine the correlations of common quantitative admissions statistics with graduate physics GPA, as well as the role that graduate GPA plays in predicting whether a student completes their PhD program.  Whereas \cite{miller_typical_2019} did not utilize information on student graduate course performance, this paper incorporates graduate GPA into several models in order to determine whether commonly used admissions metrics predict PhD completion of US students directly, or indirectly via graduate GPA; a discussion of the theoretical motivation for why graduate grades may mediate the relationship between admissions metrics and PhD completion is offered in Section \ref{sec:Background}.  Hence, while the analysis presented in \citeauthor{miller_typical_2019} \cite{miller_typical_2019} was primarily focused on simply identifying the measures that best correlated with PhD completion, this analysis explores questions regarding both how and why those correlations occurred. 

Exploring whether graduate GPA mediates the relationship between common admissions metrics and PhD completion affords us the opportunity to employ statistical methods from the literature on causal inference \cite{robins_identifiability_1992, pearl_causality_2000, pearl_direct_2001, vanderweele_explanation_2015, hayes_introduction_2013, pearl_causal_2012, imai_general_2010, valeri_mediation_2013}.  In doing so we lay out methods of calculating the direct and indirect effects of common admissions metrics on PhD completion, as well as the assumptions needed for those effects to have a causal interpretation.  This approach allows us to gain useful information from the present analysis, while careful examination of the assumptions required for causal interpretation will help guide future studies.  

Use of statistical methods developed in the causal inference literature allows us to build on the findings in \cite{miller_response} by incorporating the ranking of a student's PhD program along a mediating pathway to completion rather than as a covariate in regression analysis.  We also present models with various combinations of GRE-P and GRE-Q scores to show that variance inflation due to collinearity is minimal, and is therefore not a concern.  These analyses are included in the Supplemental Material.


We seek to answer two primary research questions in this paper:
\begin{enumerate}
    \item How do commonly used admissions metrics and demographic factors relate to physics graduate GPA?
    \item What role does graduate GPA play in predicting PhD completion, and do quantitative admissions metrics predict PhD completion indirectly through graduate GPA?
\end{enumerate}

To answer these questions, we begin by exploring the relationships between variables using bivariate correlations.  We then examine the unique predictive effects of different admission metrics on graduate GPA using a multiple linear regression model.  These results lay the groundwork for a mediation analysis, which is used to examine the role that graduate GPA plays in PhD completion by breaking down effects into direct and indirect components.  All of the primary analyses are performed using data on US physics graduate students, with a review of equivalent analyses for international students included in the Supplemental Materials.

\section{\label{sec:Background} Background and Motivation}
Before outlining the quantitative methods employed in this analysis, we briefly describe the student performance metrics used in this study and the broad individual student characteristics they help to measure.  We discuss the underlying constructs hypothesized by the GRE Quantitative, Verbal, and subject tests, as well as undergraduate and graduate grades, and several external factors that influence these scores.  The GRE Analytical Writing test is not included since it is not used enough in physics graduate admissions to warrant investigation.  This section serves as a theoretical motivation for the models of PhD completion analyzed in this study.

The GRE is a series of standardized tests designed to help admissions committees predict future academic success of students coming from different backgrounds \cite{ets2020guide, wendler_research_nodate}.  While the GRE-Q assesses basic concepts of arithmetic, algebra, geometry, and data analysis, the GRE-V assesses reading comprehension skills and verbal and analytical reasoning skills.  These tests are specifically constructed to measure ``basic developed abilities relevant to performance in graduate studies" \cite{briel1993manual}.  In their meta-analysis of GRE predictive validity, \citeauthor{kuncel2001comprehensive} frame the GRE-Q and GRE-V as most related to declarative and procedural knowledge and suggest that they are best described as measures of general cognitive ability \cite{kuncel2001comprehensive}.  In contrast, the GRE subject tests ``assess acquired knowledge specific to a field of study" \cite{ets2020guide}, indicating that the GRE subject tests are ostensibly a direct measure of a student's knowledge of a particular area of study.  Indeed, admissions committees often interpret high GRE subject scores as strong evidence of a student's discipline-specific knowledge \cite{walpole_selecting_2002}.  Other research suggests that higher scores on standardized subject tests could also reflect greater student interest in that subject area \cite{willingham2002grades}.   

The individual characteristics measured by a student's undergraduate grades include both academic knowledge and a collection of noncognitive factors \cite{bowers_what_2016}.  Much research exists on the meaning and value of grades, particularly at the K-12 level, and a review \cite{brookhart_century_2016} of the past century of grading research finds that grades assess a multidimensional construct comprising academic knowledge, engagement (including motivation and interest), and persistence.  Consistently over the past 100 years only about 25\% of variance in grades is attributable to academic knowledge as measured by standardized tests \cite{bowers_whats_2011}, with recent research suggesting that much of the unexplained variance is represented by a student’s ability to negotiate the ``social processes" of school \cite{bowers_reconsidering_2009}.   We therefore regard UGPA as broadly measuring student academic achievement across a wide range of subjects in addition to several aspects of noncognitive traits such as motivation, interest, and work habits.  However, we also recognize the limitations inherent in compressing students' college academic performance into a single number, including the loss of information pertaining to student growth over time and time to degree completion.

We conceptualize graduate GPA similarly, treating it as a measure of subject-specific academic knowledge as well as other non-academic characteristics.  In addition to the broad research on grades described above, research specifically addressing the factors leading to graduate success supports this interpretation of GGPA.  Interviews conducted with over 100 graduate school faculty reveal that graduate success, which they define as a student's ability to earn high graduate grades and eventually complete their degree in a timely manner, is largely dependent on noncognitive characteristics \cite{walpole_selecting_2002}.  Interviewees deemed motivation, work ethic, maturity, and organizational skills as crucial to student success in graduate school.  In a separate review of noncognitive predictors pertaining to graduate success, graduate GPA is specifically linked to a variety of personality (e.g. extroversion and conscientiousness) and attitudinal factors (e.g. motivation, self-efficacy, and interests) \cite{kyllonen_role_2011}. Indeed, the authors of the review characterize graduate grades as a complex composite of many of the cognitive and noncognitive factors related to graduate school success.   

These conceptions of grades and GRE scores compel us to hypothesize that graduate GPA mediates the relationship between common quantitative admissions metrics and PhD completion.  As a construct measuring subject-specific knowledge and several noncognitive characteristics, we expect UGPA and GRE-P to most strongly link to GGPA. Despite the drawbacks of cumulative UGPA (such as grade inflation and masking of individual growth), we expect UGPA to be associated with graduate course performance since it captures aspects of both academic and some non-academic characteristics.  We also expect GRE-P scores to be related to graduate grades due to their requirement of specific physics knowledge.  Finally, while we expect GRE-Q may have a small predictive effect on GGPA as a general cognitive measure, we do not necessarily expect a similar relationship for GRE-V as its content is generally disparate from physics curricula.  

On both a theoretical and structural level we expect graduate grades to predict PhD completion.  Graduate GPA may offer insight into a student's mastery of advanced physics concepts as well as their personality and attitudes.  All of these contribute to successful physics PhD completion, but likely vary in importance depending on choice of research area \cite{kyllonen_role_2011}.  Structurally, a satisfactory performance in graduate courses is implicit on the path to completing a PhD.  For example, GGPA requirements can act as thresholds for being allowed to continue studying in a PhD program.  Poor course performance may negatively influence personal factors (e.g., self-efficacy, identity), limit access to research opportunities (e.g., repeating classes, ease of finding a research lab), or may indicate a lack of preparation for research, all of which could hinder PhD completion.  It is also more temporally proximal to PhD completion than the other metrics included in the study.

Lastly, we note that although this discussion has focused on students' individual traits that may predict success in graduate school, there are undoubtedly a number of external factors that can influence student attrition.  Socioeconomic factors, mental health, family responsibilities, work duties, external job prospects, and departmental culture are all variables that would play a role in a comprehensive model of graduate school persistence \cite{walpole_selecting_2002, owens_student_nodate, owens2018, gardner2008fitting, lovitts2002leaving, national2018graduate}.  These uncollected pieces of information may act as ``confounding" variables that can bias results, and we discuss their influence on the present study in Section \ref{subsec:Limitations}. 

\section{\label{sec:Method}Method}
\subsection{\label{subsec:Data}Data}

\begin{figure*}
    \includegraphics[]{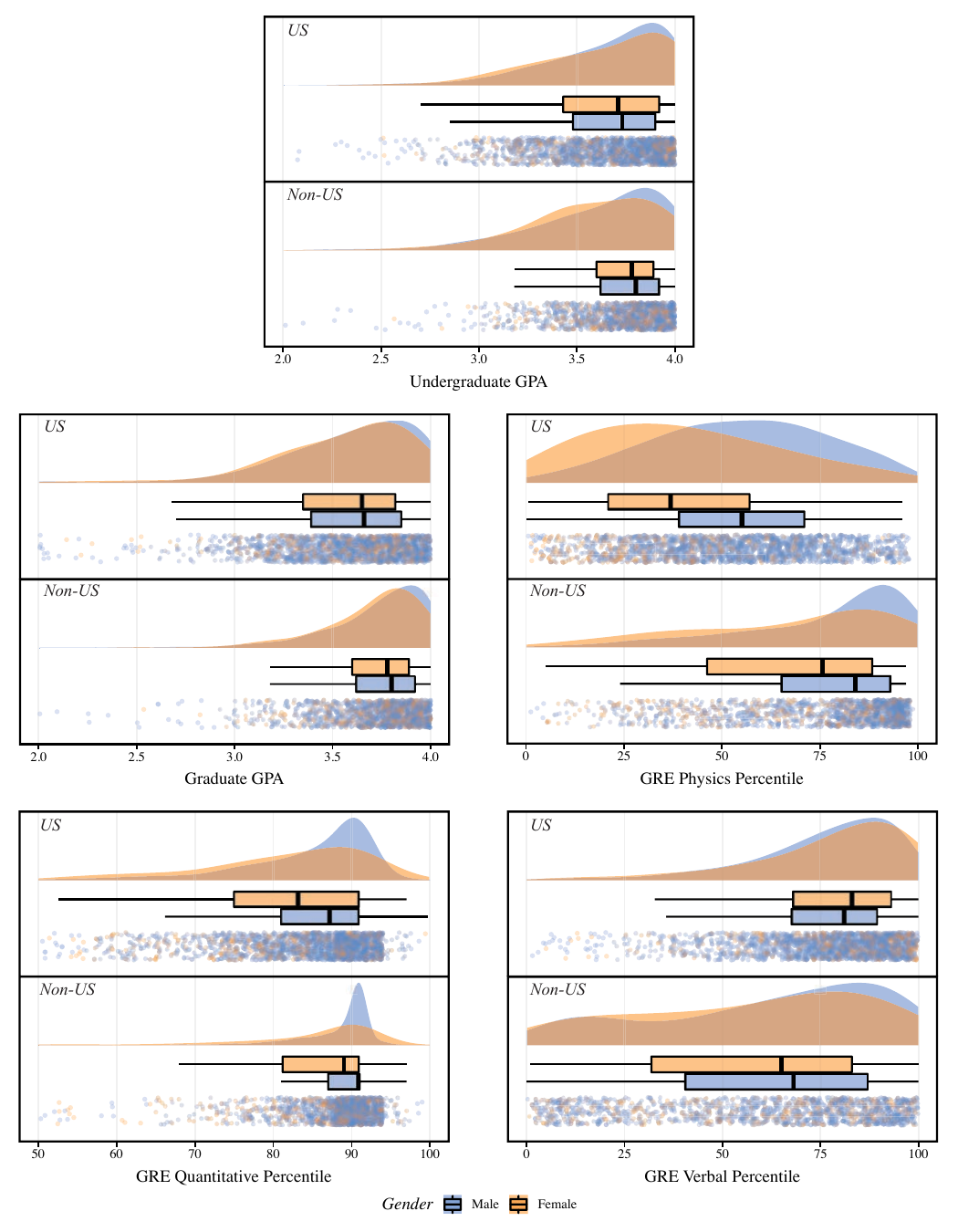}
    \caption{\label{fig:distributions} Distributions of the quantitative metrics included in the data. ``Raincloud plots" show density plots, boxplots, and scatterplots of the data. We see that despite significant score gaps between US male and female GRE-P test takers, no such gap exists in subsequent GGPA performance.  UGPA distributions for male and female applicants are also similar. Code for generating raincloud plots courtesy of \cite{allen_raincloud_2019}. All figures generated with the R package \texttt{ggplot2} \cite{ggplot_package}. Figure themes adapted from \cite{desiraju_rpubs_nodate}.}
\end{figure*}



Student level data for both this study and \cite{miller_typical_2019} was requested from physics departments that awarded more than 10 PhD's per year for students who matriculated between 2000 and 2010, including information on undergraduate GPA (UGPA), GRE-Q, GRE-V, GRE-P, and graduate GPA (GGPA).  Data collected also included the final disposition of students (PhD earned or not), start and finish years, and demographic information. GPA data is analyzed on a 4.0 scale while GRE scores are on the percentile scale.

We received data from 27 programs (approximately a 42\% response rate), which spanned a broad range of National Research Council (NRC) rankings.  The sample used in \cite{miller_typical_2019} consisted of all students in 21 programs for which start year was available.  Given that the median time to degree across physics PhD programs is 6 years, some students who started before 2010 were still active at the time of data collection in 2016. The probability of not completing the physics PhD has an exponential time dependence with a time constant of 1.8 years. Thus, students who have been in their programs for three time constants have only a 5\% chance of not completing. These students were thus categorized as completers in this study.  

These data covered 3962 students.  Of this subset, two programs did not report GGPA data for their students.  Hence, the sample for this study excludes these students, thereby reducing the sample size to 3406 students across 19 programs. This corresponds to approximately 11\% of matriculants to all U.S. physics PhD programs during the years studied.

\begin{table}[t]
\def\arraystretch{1.3}
\begin{tabularx}{\columnwidth}{Y|Y|Y|Y|}
\cline{2-4} & US   & Non-US & Total \\ \hline
\multicolumn{1}{|Y|}{Male}   & 1638 & 1164   & 2802  \\ \hline
\multicolumn{1}{|Y|}{Female} & 317  & 287    & 604   \\ \hline
\multicolumn{1}{|Y|}{Total}  & 1955 & 1451   & 3406  \\ \hline
\end{tabularx}
\caption{\label{tab:demographics} Demographic breakdown of the data used in this analysis.  To focus on issues of diversity and inclusion most strongly associated with US applicants to physics graduate programs, we analyze only the data from US graduate students.}
\end{table}
Among the sample of US students, 16\% are women ($N=317$).  Although the authors generally advocate for a nuanced treatment of gender in physics education research and recognize the deficits associated with treating gender as a fixed binary variable \cite{traxler_enriching_2016}, the present data set spans the years 2000 to 2010 during which the data collected by programs only allowed for the binary option of male/female.  Hence, we must treat gender as a dichotomous variable in this analysis.  

The racial composition of the dataset is 61.6\% White, 1.3\% Black, 2.1\% Hispanic, 0.2\% Native American, 3.5\% Asian, 1.0\% multiple or other races, and 30.2\% undisclosed.  Excluding the cases for which race was unavailable, the sample is thus roughly representative of annual PhD production in U.S. physics for gender, race, and citizenship \cite{noauthor_deptofed_nodate}.  We include race as a covariate in each analysis presented; however small $N$, particularly for Black, Hispanic, Native American, and Asian students, often precludes useful interpretation of the results pertaining to these subsets.   

In order to focus on issues of diversity and inclusion associated most strongly with US applicants, we use only the subset of data from domestic graduate students.  This decision is further motivated by research suggesting that it is difficult for admission committees to directly compare scores earned by US and international students, indicating that separate analyses are appropriate \cite{walpole_selecting_2002}.  Using the subset of students who are from the US reduces the total sample size for the study to $N=1955$. A cursory visualization of the variables in the data set, as shown in Figure \ref{fig:distributions}, shows that the distributions of scores for US and Non-US students are markedly different, which further justifies separate analyses of these two student populations.  

Examining the distributions of scores in Figure \ref{fig:distributions}, the presence of non-normality is evident in nearly all of the variables.  Each of the continuous variables in the dataset fail the Shapiro-Wilk test of normality at the $\alpha$ level of .05.  However, these tests are often of limited usefulness; in general distributions with skewness $\lvert \hat{\gamma_1} \rvert >3$ or kurtosis $\lvert \hat{\gamma_2} \rvert >10$ likely indicate that they violate any assumption of normality \cite{kline_principles_2015}.  For this dataset, the GGPA distribution skewness $\hat{\gamma_1} = -3.31$ and kurtosis $\hat{\gamma_2} = 20.43$, indicating severe non-normality.  The GRE-Q distribution also falls into the problematic range ($\hat{\gamma_1} = -2.11$ and $\hat{\gamma_2 = 9.33}$).  Ceiling effects are also present, since many students earned 4.0 grade point averages or earned the maximum score on the GRE examinations.

The data collection process was limited to gathering only cumulative graduate GPA rather than first-year graduate grades, which were not recorded by some programs.  Thus, depending on whether a student persisted in a program, their graduate GPA may be based on many courses while others are based on only a few courses.  The data set is also necessarily subject to range restriction, since data on student performance in graduate school is automatically limited to include only students who were accepted to undertake graduate study.  We cannot know how students who were not accepted into graduate school would have performed had they been accepted.  Range restriction may act to attenuate the strength of observed effects in subsequent analyses \cite{small2017range}.     

Although not used in a majority of this study, we briefly explore the role of the doctoral programs’ NRC ranking in PhD completion \cite{national2010data}.  Since the NRC only gives confidence intervals for program rank, we created a ranking for this study by averaging the 5 and 95\% confidence bounds for the NRC regression-based ranking (NRC-R) and rounded this up to the nearest five to protect the confidentiality of participating programs. This led to a ranking range of 5 to 105. We divided the programs into terciles of approximately equal number of records, and categorized as Tier 1 (highest ranked, NRC-R $\leq$ 20), Tier 2 (25 $\leq$ NRC-R $\leq$ 55), and Tier 3 (NRC-R $>$ 55).

Multiple imputation (predictive mean matching) is used to impute missing UGPA and GRE-P scores.  Predictive mean matching is used due to the non-normality of the data.  160 students do not have data for either UGPA or GRE-P, while 400 are missing only UGPA and 263 are missing only GRE-P.  All multiple imputation is conducted using the \texttt{mice} package in R \cite{mice_package}.  20 imputed data sets are used for each analysis.  For consistency, incomplete variables are imputed using the same imputation model used in \cite{miller_typical_2019}, in which the imputation model utilizes all other variables in the data set aside from graduate GPA and PhD completion.  The model utilizes GRE-Q, GRE-V, program tier, gender, and race, as well as complete cases of UGPA and GRE-P.  Although the imputation approach presented here is theoretically sound, we also present a comparison of several different models of data imputation in the Supplemental Materials.

\subsection{\label{subsec:GraduateOutcome} Methods to explore the role of graduate grades}

The goal of this section is twofold.  First, we seek to gain a cursory look at how graduate GPA is related to common admissions metrics.  In doing this, we also wish to determine whether it is reasonable that admissions metrics could indirectly predict completion through graduate GPA.  This section presents a series of analyses meant to elucidate the relationships between standard admissions metrics, students' GGPA, and students' final disposition.  To make our analysis maximally accessible to readers of different statistical backgrounds, we describe here in detail the methods used in this section.

%
%

Bivariate correlation coefficients provide information about the level of association between two variables, and are therefore a useful starting point for analysis.  We construct a correlation matrix (see Table \ref{tab:correlations}) for all variables in the sample using Pearson correlation coefficients, which are equivalent to the standardized slope coefficients for a linear model predicting $y$ from $x$.  These are given by
\begin{equation}
    r_{xy} = \frac{\sum (x_i-\mean{x})(y_i-\mean{y})}{\sqrt{\sum (x_i-\mean{x})^2} \sqrt{\sum (y_i-\mean{y})^2}}
    \label{eq:pearson}
\end{equation}
for any two continuous variables $x$ and $y$.  Calculating $r_{xy}$ gives us a first glance at the relationships between the continuous variables UGPA, GGPA, GRE-P, GRE-Q, and GRE-V. 

When the $x$ variable is treated as dichotomous (e.g., gender in this data set), Eq. (\ref{eq:pearson}) reduces to the point-biserial correlation coefficient $r_{pb}$,
\begin{equation}
    r_{pb} = \frac{\mean{y_1}-\mean{y_0}}{\sigma_y}\sqrt{pq},
\end{equation}
where $\mean{y_1}$ and $\mean{y_0}$ are the means of the continuous $y$ variable for the two $x$ groups 1 and 0, $q$ and $p$ are the proportions of data belonging to these two groups, and $\sigma_y$ is the standard deviation for the $y$ variable.  Like the Pearson coefficient, the quantity $r_{pb}$ ranges from -1 to 1 and indicates the strength of association between two variables.  Conveniently, a significance test for the point-biserial correlation is identical to performing an independent \emph{t}-test on the data \cite{field_discovering_2012}.  Thus, the point-biserial correlation coefficient yields information about whether two group means are statistically different.  For instance, the point-biserial correlation tests whether the GGPA of male students are statistically different from those of female students (we find that GGPAs are not significantly different by gender, see Table 
\ref{tab:correlations}).

When $x$ and $y$ are both dichotomous, the Pearson coefficient reduces to the phi coefficient,
\begin{equation}
    \phi = \sqrt{\frac{\chi^2}{n}},
\end{equation}
where $\chi^2$ is the chi-squared statistic for a 2x2 contingency table and $n$ is the total number of observations in the data.  The phi coefficient also ranges from -1 to 1 and indicates the strength of association between two binary variables.  This quantity allows us to examine whether final disposition is significantly associated with gender (we find that the association between gender and final disposition just meets the threshold for statistical significance ($\phi=-0.05 \pm 0.04$, $p=0.04$), likely due to the large sample size of our data, but the very small phi coefficient indicates that the practical strength of this relationship is negligible \cite{scott_jones_learn_2021}).   

%
%

To characterize how GGPA and other numerical predictors vary across program tier, we conduct several one-way analysis of variance (ANOVA) tests using program tier as the independent variable.  ANOVA tests allow us to determine whether there are significant differences between different groups, such as students in different program tiers.  These tests produce an \emph{F}-statistic, which is interpreted as the ratio of between-group variability to within-group variability.  Thus, higher values of \emph{F} indicate that between-group variability is large compared to within-group variability, which is unlikely if the group means all have a similar value.  

%
%

Lastly we present the results of a multiple regression analysis in which we regress GGPA on common admissions metrics and demographic factors.  Regression allows us to examine the unique predictive effects of these predictors.  

The classical linear regression model is written mathematically for an outcome variable $Y$ as 
\begin{equation}
    Y_i = \alpha_1 X_{i1} + \alpha_2 X_{i2} + ... + \alpha_k X_{ik} + \epsilon_i,
\end{equation}
where $i = 1,...,n$, the number of observations in the data, and $k$ represents the number of predictors in the model.  Error terms $\epsilon_i$ are assumed to be independent and normally distributed with mean 0 and standard deviation $\sigma$.  $\hat{\alpha}$ is the vector of regression coefficients that minimizes the sum of squared errors
\begin{equation}
    \Sigma_{i=1}^n = (Y_i - \hat{\alpha}X_i)^2
\end{equation}
for the given data.  The regression coefficients can be interpreted as the difference in the outcome variable $Y$, on average, when comparing two groups of units that differ by 1 in one predictor $X$ while keeping all the other predictors the same.  

We report both unstandardized and standardized versions of the regression coefficients.  Unstandardized coefficients are the result of regression analysis using the original, unscaled variables.  Thus, the unstandardized regression coefficients represent the predicted average change in the outcome variable $Y$ when the corresponding predictor $X$ is changed by one unit.  This allows for a straightforward interpretation since the variables are not scaled, but does not yield insight into the relative predictive strengths of the independent variables since they are scaled differently.  Standardized regression coefficients result from regression analyses using continuous variables that have been mean-centered and divided by their standard deviation, resulting in variables with variances equal to 1.  Thus standardized regression coefficients represent the average number of standard deviations changed in the outcome variable when a predictor variable is increased by one standard deviation.  By calculating the standardized coefficients, we exchange a simple interpretation of score change for an interpretation of which variables have the greatest effect on the dependent variable.

\subsection{\label{subsec:Mediation} Mediation analysis methods}

Using mediation analysis we seek to answer the question of whether graduate GPA mediates the predictive ability of common admissions metrics on PhD completion.  Whereas analyses such as logistic regression \cite{miller_typical_2019} yield information about whether independent variables such as UGPA and GRE-P affect final disposition of a graduate student, they do not offer insight into the explanation of why and how UGPA, GRE-P, and other admissions metrics affect completion. Mediation analysis is one technique that allows us to probe the underlying process by which some variables influence others \cite{vanderweele_explanation_2015, hayes_introduction_2013, pearl_causal_2012, imai_general_2010, valeri_mediation_2013}.  

\begin{figure}
    \includegraphics[]{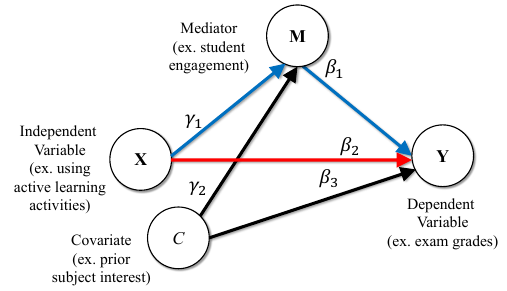}
    \caption{\label{fig:mediation} A qualitative graphical depiction of a mediated relationship between two variables. $X$, $Y$ and $M$ represent the model's independent, dependent, and mediating variables while $C$ represents a covariate.  A researcher who observes a positive relationship between using active learning activities in the classroom and the exam grades of students might posit that a third variable, student engagement, is actually responsible for causing the observed relationship.  }
\end{figure}

Figure \ref{fig:mediation} graphically depicts a prototypical mediation model, where $X$, $Y$ and $M$ represent the model's independent, dependent, and mediating variables.  $C$ represents a covariate.  As a hypothetical example, let's say that previous research has shown a positive relationship between the use of active learning activities in physics class and student exam grades.  Researchers might hypothesize that this relationship is actually due to a third mediating variable, student engagement. Using active learning activities in class may cause students to become more engaged with the material, making their subsequent exam grades increase. Engagement is a mediating variable in this case.  Meanwhile, since students who are already interested in physics could be predisposed to being more engaged and performing better on exams, the researcher might take students' prior physics interest into account by including it as a covariate in their analyses. 

In this section, we wish to discern whether graduate GPA mediates the relationship between common admissions metrics and students' likelihood of completing graduate school.  In practice, mediation analysis is done by simultaneously estimating a set of regression equations \cite{muthen_regression_2016}.  The goal is to partition the total effect of the independent variable $X$ on the dependent variable $Y$ into two parts: the direct effect of $X$ on $Y$ and the indirect effect of $X$ on $Y$ through the mediating variable $M$.  For the simple example given above, the set of regression equations to be solved are:
\begin{equation}
    Y_i = \beta_0 + \beta_1 M_i + \beta_2 X_i + \beta_3 C_i + \epsilon_{yi}
    \label{eq:Y_mediation}
\end{equation}
\begin{equation}
    M_i = \gamma_0 + \gamma_1 X_i + \gamma_2 C_i + \epsilon_{mi}.
    \label{eq:M_mediation}
\end{equation}
Traditional mediation literature \cite{baron_moderatormediator_1986, mackinnon_introduction_2008} defines the direct effect of $X$ on $Y$ as the coefficient $\beta_2$ and the indirect effect of $X$ on $Y$ as the product of coefficients $\gamma_1\beta_1$, corresponding to the products of the path coefficients along the mediated path shown in Figure \ref{fig:mediation}.  Statistically significant values of $\gamma_1\beta_1$ indicate that the relationship between $X$ and $Y$ is mediated by $M$. This method demonstrates the general intuitive ideas underlying mediation analysis, but is subject to several important limitations.  Foremost among the limitations associated with traditional mediation analysis is that its applicability to model categorical variables (e.g., binary outcomes) and nonlinearities is not well defined, as these situations preclude the use of sums and products of coefficients \cite{mackinnon_intermediate_2007, vanderweele_explanation_2015, hoyle_handbook_2012}.  The difficulties associated with binary outcomes are therefore problematic for a model predicting final disposition, a binary outcome.  Furthermore, traditional mediation models leave the causal interpretation of their results ambiguous \cite{sobel_identification_2008}.  

Recent work in the field of causal inference \cite{robins_identifiability_1992, pearl_causality_2000, pearl_direct_2001} has formalized and generalized mediation analysis to resolve these limitations, allowing for categorical outcomes while also clarifying that under certain conditions the results may be interpreted causally.  In this framework, often called the ``potential outcomes" framework, the traditional product-of-coefficients mediation analysis is a special case for which the mediator and outcome variables are both continuous, while the functional forms of the direct and indirect effects for other situations become more complicated \cite{muthen_regression_2016}.

For the primary analysis of this paper, we calculate the direct and indirect effects defined by the potential outcomes framework for the case of a continuous independent variable (UGPA, GRE-P, and GRE-Q), a continuous mediator (GGPA), and a dichotomous outcome (final disposition).  The simultaneous regression equations to be calculated are still Eqs. (\ref{eq:Y_mediation}) and (\ref{eq:M_mediation}), except the binary $Y$ is replaced with $Y^*$, a continuous unobserved latent variable which represents the observed binary variable.  Once estimated, the direct and indirect effects reduce to simple differences in probability of completing a PhD between students across different values of the independent variables.  Mathematically, the effects for a change in the independent variable from a value $x_0$ to $x_1$ at a particular value of the control $c$ are given by
\begin{equation}
    \text{IE} = \Phi[\text{probit}(x_0,x_1)] - \Phi[\text{probit}(x_0,x_0)],
    \label{eq:indirecteffect}
\end{equation}
\begin{equation}
    \text{DE} = \Phi[\text{probit}(x_1,x_1)] - \Phi[\text{probit}(x_0,x_1)],
    \label{eq:directeffect}
\end{equation}
where $\Phi$ represents the normal cumulative distribution function and $\text{probit}(x_a,x_b)$ is given by
\begin{multline}
    \text{probit}(x_a,x_b) = [\beta_0 + \beta_2x_a + \beta_3c + \\
    \beta_1(\gamma_0 + \gamma_1x_b + \gamma_2c)]/\sqrt{v(x_a)},
\end{multline}
and $v(x_a)$ is
\begin{equation}
    v(x_a) = \beta_1^2 \sigma_m^2 + 1.
\end{equation}
Note that these expressions are all still simply combinations of the coefficients from the regression equations (\ref{eq:Y_mediation}) and (\ref{eq:M_mediation}). Thus, the potential outcomes framework allows us to calculate the total predicted change in probability of completing a PhD due to an independent variable and decompose it into that variable's indirect effect on final disposition through GGPA as well as its direct effect (see Figure \ref{fig:mediations}).  

Using this mediation framework can help to give powerful insights into nuanced relationships between the variables in an observational study \cite{imai_unpacking_2011}.  However, giving a truly causal interpretation to the results of this mediation analysis requires a set of strong assumptions to be met, and in practice it can be difficult for any observational study fully meet these conditions \cite{muthen_applications_nodate}.  Hence, we try to avoid making explicitly causal claims in our discussion of the results.  We discuss the assumptions needed for a causal interpretation as well as the robustness of the current study to violations of those assumptions in Section \ref{subsec:Limitations}.

Mediation analyses were conducted using the \texttt{mediation} package in R \cite{mediation_package}.  Checks for the consistency of results across different computational approaches were done by performing duplicate mediation analyses in the R package \texttt{medflex} \cite{medflex_package} as well as the statistical software Mplus \cite{muthen2009mplus}.

\begin{figure}
\includegraphics[]{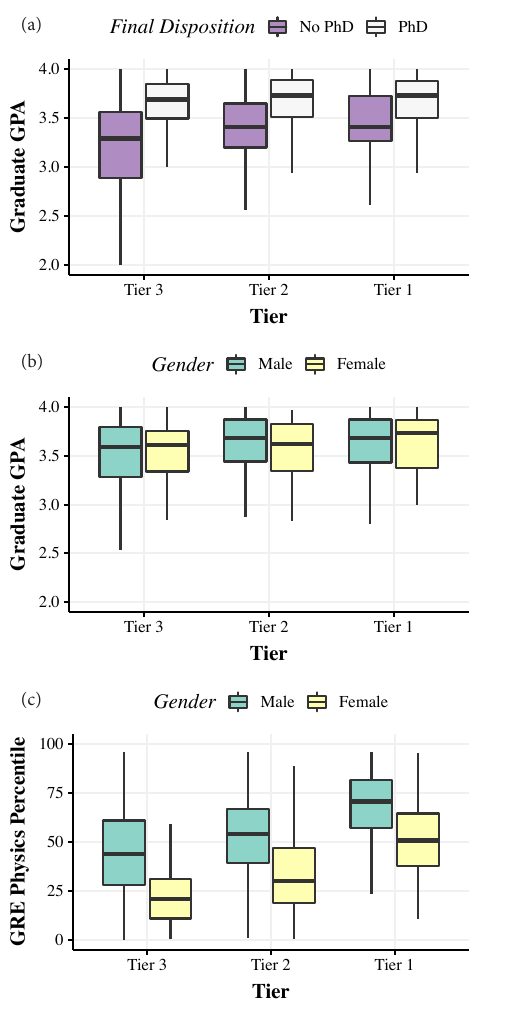}
\caption{\label{fig:boxplots} (a) Graduate GPA by program tier and final disposition. Students who do not complete a PhD earn lower graduate grades than students who complete their programs ($r_{pb}=0.43$, $p<.001$). Tier 3 students tend to earn slightly lower graduate course grades than Tier 1 or 2 students. (b) Graduate GPA by program tier and gender. Male and female students earn similar graduate grades ($r_{pb}=-0.01$, $p=0.57$, equivalent to a non-significant independent $t$-test), and this trend holds across program tier. (c) GRE Physics by program tier and gender. Across all program tiers there is a significant gap in scores between male and female GRE-P test takers ($r_{pb}=-0.30$, $p<.001$, equivalent to a statistically significant independent $t$-test). Score distributions trend upward for higher tier programs.}
\end{figure}

\begin{table*}[t]
\centering
\def\arraystretch{1.2}
\begin{tabular}{lccccccc}
  \hline
 Measure (M + SD) & UGPA & GRE-Q & GRE-V & GRE-P & GGPA & Final Disp. & Gender \\ 
  \hline
  UGPA (3.6 $\pm$ 0.3) & -- & (0.25, 0.35)  & (0.12, 0.22)  & (0.26, 0.37)  & (0.24, 0.33)  & (0.10, 0.20)  & (-0.11, 0.01)  \\ 
  
  GRE-Q (83.3 $\pm$ 10.4) & 0.30 & -- & (0.30, 0.37)  & (0.47, 0.54)  & (0.13, 0.22)  & (0.10, 0.18)  & (-0.16, -0.07)  \\ 
  
  GRE-V (76.3 $\pm$ 18.7) & 0.17 & 0.33 & -- & (0.23, 0.32)  & (0.06, 0.15)  & (0.02, 0.10)  & (-0.02, 0.07)  \\ 
  
  GRE-P (52.9 $\pm$ 23.2) & 0.31 & 0.51 & 0.28 & -- & (0.18, 0.27)  & (0.10, 0.19)  & (-0.33, -0.24)  \\ 
  
  GGPA (3.5 $\pm$ 0.5) & 0.29 & 0.18 & 0.10 & 0.22 & -- & (0.39, 0.46)  & (-0.06, 0.03)  \\ 
  
  Final Disp. & 0.15 & 0.15 & 0.06 & 0.14 & 0.43 & -- & (-0.09, -0.01)  \\
  
  Gender & -0.05 & -0.11 & 0.03 & -0.30 & -0.01 & -0.05 & --  \\ 
  
   \hline
\end{tabular}
\caption{\label{tab:correlations}A matrix showing bivariate correlations between continuous and dichotomous variables used in subsequent analyses.  Correlations are shown in the lower diagonal while confidence intervals for those correlations are shown in the upper diagonal.  For example, the correlation between GGPA and GRE-P is $0.22$ and a 95\% CI of (0.18, 0.27), indicating a weak correlation. Means and standard deviations are also presented in the first column.  GPAs are on a 4.0 scale while GRE scores are in terms of percentiles.  Correlations are calculated for US students only.}
\end{table*}

\section{\label{sec:Results}Results}

\subsection{\label{subsec:RoleofGGPA}Results of exploring the role of graduate grades}

%
%

\textbf{Correlations} An initial question related to predicting a student's final disposition is whether UGPA, GRE scores, and GGPA are reliably correlated with a student's final outcome.  We are also interested in the strength of association between GGPA, UGPA, and GRE scores, as this information yields insight into whether GGPA could serve as a mediating variable in predicting final disposition.  Table \ref{tab:correlations} contains the bivariate correlations (Pearson, point-biserial, and phi) between each pair of measures for the sample in the lower diagonal.  The 95\% confidence intervals are reported in the upper diagonal.  Confidence intervals that do not include a value of zero indicate that the correlation is statistically significant.  The means and standard deviations of the continuous variables are presented in the table's first column (GPA data is analyzed on a 4.0 scale while GRE scores are on the percentile scale).

Inspection of Table \ref{tab:correlations} reveals that GGPA is the predictor most strongly correlated with final disposition ($r_{pb}=0.43$).  This value is statistically significant ($p<.001$) and positive, meaning that students with higher GGPA are more likely to finish their PhD program successfully.  This trend is visually apparent in Figure \ref{fig:boxplots}(a), which shows boxplots of GGPA grouped by PhD completers and non-completers.  We also observe that UGPA ($r_{xy}=0.29$, $p<.001$), GRE-Q ($r_{xy}=0.18$, $p<.001$), and GRE-P ($r_{xy}=0.22$, $p<.001$) are all positively correlated with GGPA, albeit weakly, meaning that students with higher scores in these metrics tend to earn higher GGPAs.  Taken together, the observation that higher UGPA and GRE scores positively correlate to GGPA, which in turn correlate with a student's likelihood of completion, implies that GGPA might play an important role in mediating the influence of these admissions metrics on PhD completion.

The lack of a statistically significant correlation between gender and GGPA indicates that the disparity in scores on the GRE-P between males and females does not manifest itself in subsequent GGPA performance.  Indeed, there is no statistical difference between average GGPA for males and females ($r_{pb}=-0.01$, $p=0.57$, equivalent to a non-significant independent $t$-test), as demonstrated in Figure \ref{fig:boxplots}(b).  Yet there exists a statistically significant difference between males and females in GRE-P performance in our data ($r_{pb}=-0.30$, $p<.001$, equivalent to a statistically significant independent $t$-test).  Thus GGPA does not differ between genders despite the known performance gap between males and females on the GRE-P exam.  Furthermore, the phi coefficient measuring the association between gender and PhD completion is negligible despite barely meeting the threshold of statistical significance ($\phi=-0.05$, $p=0.04$).

Still, the bivariate correlations shown in Table \ref{tab:correlations} do not control for possible relationships between the variables of interest.  For instance, the moderate correlation between GRE-Q and GRE-P ($r_{xy}=0.51$, $p<.001$) indicates that there may be a spurious relationship between one of these variables and GGPA.  In addition, we observe low but statistically significant correlations between UGPA and GRE-P ($r_{xy}=0.31$, $p<.001$), GRE-Q ($r_{xy}=0.30$, $p<.001$), and GRE-V ($r_{xy}=0.17$, $p<.001$).  This is expected, as UGPA likely contains some information regarding the specific aspects of students' aptitudes tested by GRE exams.  These results motivate the use of multiple regression analysis later in this section in order to disentangle the unique effects of each independent variable on GGPA.  That analysis reveals that when we isolate the unique predictive effects of each variable in the regression model, UGPA and GRE-P remain significant but weak predictors of GGPA, while GRE-Q does not retain statistical significance. 

Similarly, although UGPA ($r_{pb}=0.15$, $p<.001$), GRE-P ($r_{pb}=0.14$, $p<.001$) and GRE-Q ($r_{pb}=0.15$, $p<.001$) are positively correlated with PhD completion, the magnitudes of these correlations are very weak and do not account for other parameters that may be associated with completion.  Multivariate approaches allows us to isolate how individual metrics relate to PhD completion, which we explore in the mediation analysis presented in Section \ref{subsec:MediationResults}.  Consistent with previous studies of PhD completion using multivariate approaches \cite{miller_typical_2019}, we find that when accounting for other parameters, only UGPA remains a statistically significant predictor of completion.

%
%
 
Results of one-way independent ANOVA tests show that the main effect of program tier on GGPA is significant, $F(2, 1949) = 26.31$, $p<.001$, which reflects the upward trend in GGPA from Tier 3 to Tier 1 and 2 programs.  A Tukey post hoc test reveals that the GGPA was significantly higher for students at Tier 2 ($M=3.59$, $SD=0.40$, $p<.001$) and Tier 1 ($M=3.60$, $SD=0.42$, $p<.001$) institutions than those at Tier 3 institutions ($M=3.41$, $SD=0.59$). There was no statistically significant difference between the Tier 1 and Tier 2 groups ($p=0.97$).  


%
%

\textbf{Multiple Regression} To disentangle the unique effects of each predictor on GGPA we conduct a multiple linear regression analysis.  Multiple linear regression allows us to simultaneously fit many independent variables to measure each of their relative effects on a single dependent variable, GGPA.  Analyzing the raw coefficients fitted by the regression analysis yields insight into the predicted change in GGPA due to changes in one variable while holding all others constant.  Standardized coefficients allow for a comparison of the relative effect sizes of the independent variables. 

Our model includes all available GRE scores (GRE-P, GRE-Q, and GRE-V) as well as UGPA in order to examine the unique predictive effects of each measure.  We considered the possibility that including both GRE-P and GRE-Q in the same model would raise collinearity concerns, but find these concerns unfounded.  The bivariate correlation ($r_{xy}=0.51$) between the two is not high enough to warrant genuine concern \cite{dormann_collinearity_2013, vatcheva_multicollinearity_2016}; furthermore, the variance inflation factor (VIF) for every imputed dataset's regression model was below 1.75, well below the commonly cited threshold of 10.  Hence, we deem the model posed in the study as most appropriate to answer the research questions raised in this study (further discussion regarding collinearity concerns in the data is available in the Supplemental Material). 


\begin{table}[t]
\begin{center}
\def\arraystretch{1.2}
\begin{tabular}{llcc}
\multicolumn{4}{c}{Multiple Regression Results (*$p<.05$, **$p<.01$)}                                                                       \\ \hline\hline
\multicolumn{2}{l}{}                            & \multicolumn{2}{c}{Dependent Variable - GGPA} \\ \cline{3-4} 
\multicolumn{2}{l}{\begin{tabular}[c]{@{}l@{}}Independent\\ Variable\end{tabular}} &
  \begin{tabular}[c]{@{}c@{}}Coefficient\\ (Standard Error)\end{tabular} &
  \begin{tabular}[c]{@{}c@{}}Standardized\\ Coefficient ($\beta$)\end{tabular} \\ \hline

\rowcolor[gray]{.9}[\tabcolsep]   
\multicolumn{2}{l}{Intercept} & 1.92** (0.15) & -0.06 \\

\rowcolor[gray]{.9}[\tabcolsep] 
\multicolumn{2}{l}{UGPA} & 0.35** (0.04) & 0.24** \\

\rowcolor[gray]{.9}[\tabcolsep] 
\multicolumn{2}{l}{GRE-P} & $31 \times 10^{-4}$** ($6 \times 10^{-4}$) & 0.15**          \\

\rowcolor[gray]{.9}[\tabcolsep] 
\multicolumn{2}{l}{GRE-Q} & $16 \times 10^{-4}$ ($12 \times 10^{-4}$) & 0.03              \\

\rowcolor[gray]{.9}[\tabcolsep] 
\multicolumn{2}{l}{GRE-V} & $3 \times 10^{-4}$ ($6 \times 10^{-4}$) & 0.01              \\

\multicolumn{2}{l}{Black} & -0.11 (0.09) & -0.23              \\

\multicolumn{2}{l}{Hispanic} & -0.01 (0.07) & -0.02  \\

\multicolumn{2}{l}{Native Am.} & 0.01 (0.23) & 0.03  \\

\multicolumn{2}{l}{Asian}  & -0.02 (0.06)  & -0.05  \\

\multicolumn{2}{l}{Other} & -0.24* (0.11)  & -0.51*  \\

\multicolumn{2}{l}{Undisclosed} & 0.07** (0.02)  & 0.15**  \\

\multicolumn{2}{l}{Gender} & 0.06* (0.03)  & 0.13* \\ \hline\hline

\multicolumn{2}{l}{\textit{N}} & 1955 &          \\

\multicolumn{2}{l}{Adjusted \textit{R}-Squared} & 0.11                      &             
\end{tabular}
\end{center}
\caption{\label{tab:regression} Coefficients of a multiple regression analysis modeling graduate GPA as a function of common quantitative admissions metrics.  Reference categories are White for race and Male for gender.}
\end{table}

The results obtained in our analysis are summarized in Table \ref{tab:regression}.  Significant predictive effects at the 95\% threshold were found for the numerical metrics UGPA ($\beta = 0.24$, $t=8.88$, $p<.01$) and GRE-P ($\beta = 0.15$, $t=5.02$, $p<.01$).  Students with higher UGPA and GRE-P scores therefore tend to receive higher GGPAs.  Among statistically significant predictors, the highest standardized coefficient is UGPA, which is larger than the GRE-P coefficient by approximately 50\%.  

The regression model predicts that for a 0.10 score increase in UGPA, GGPA is expected to increase on average by 0.035 points, holding all other predictors fixed.  Meanwhile, a 10 percentile increase in GRE-P score is associated with a 0.031 point increase in GGPA on average, again holding other predictors fixed.  

A significant predictive effects was found for gender ($\beta = 0.13$, $t=2.11$, $p=.04$). The positive $\beta$ coefficient indicates that if a female student and male student have the same admissions scores, the female student would earn a higher grades in her graduate classes.  We include race as a parameter in the analysis as a categorical variable, though all groups except White had small $N$ (the most students in the dataset identified as White, $N=1205$, while the fewest identified as Native American, $N=4$). None of the race categories were statistically significant parameters except ``Other."   

Notably the GRE-Q coefficient is not statistically significant, suggesting that it has little predictive effect on graduate course performance.  Yet in \cite{miller_typical_2019}, GRE-Q was seen to link to overall completion.  This observation, along with those made above regarding the predictive effects of UGPA and GRE-P on GGPA, motivate the mediation analysis in the following section.  Some predictors, namely GRE-Q, may link more directly to PhD completion, while others such as UGPA and GRE-P may indirectly predict completion through their effect on GGPA, which itself links to completion.


\subsection{\label{subsec:MediationResults}Mediation analysis results}

\begin{table}[b]
\def\arraystretch{1.2}
\begin{tabularx}{\columnwidth}{|X|Y|Y|Y|}
\hline
\multicolumn{4}{|c|}{\begin{tabular}[c]{@{}c@{}}Mediation analysis results \\for score changes from 25th to 75th percentiles\end{tabular}} \\ \hline
\begin{tabular}[c]{@{}l@{}}Independent\\ Variable\end{tabular} & \textbf{\begin{tabular}[c]{@{}c@{}}Direct\\ Effect\end{tabular}} & \textbf{\begin{tabular}[c]{@{}c@{}}Indirect\\ Effect\end{tabular}} & \textbf{\begin{tabular}[c]{@{}c@{}}Total\\ Effect\end{tabular}} \\ \hline
UGPA & \begin{tabular}[c]{@{}c@{}}-0.001\\ (-0.033, 0.034)\\ $p=0.97$\end{tabular} & \begin{tabular}[c]{@{}c@{}}0.060\\ (0.037, 0.084)\\ $p<0.01$\end{tabular} & \begin{tabular}[c]{@{}c@{}}0.060\\ (0.030, 0.090)\\ $p<0.01$\end{tabular} \\ \hline
GRE-P & \begin{tabular}[c]{@{}c@{}}-0.006\\ (-0.047, 0.035)\\ $p=0.78$\end{tabular} & \begin{tabular}[c]{@{}c@{}}0.042\\ (0.023, 0.063)\\ $p<0.01$\end{tabular} & \begin{tabular}[c]{@{}c@{}}0.037\\ (-0.002, 0.075)\\ $p=0.08$\end{tabular} \\ \hline
GRE-Q & \begin{tabular}[c]{@{}c@{}}0.024\\ (-0.009, 0.060)\\ $p=0.13$\end{tabular} & \begin{tabular}[c]{@{}c@{}}0.009\\ (-0.007, 0.023)\\ $p=0.26$\end{tabular} & \begin{tabular}[c]{@{}c@{}}0.034\\ (-0.004, 0.068)\\ $p=0.07$\end{tabular} \\ \hline
\end{tabularx}
\caption{\label{tab:exampleMediation} Results of mediation analyses that show the predicted change in probability of PhD completion due to changes in admissions metrics from their 25th percentile values to their 75th percentile values among students in the study. The predicted total effect of UGPA is to increase a student's overall probability of PhD completion by 6.0\% ($p<0.01$).  That effect is entirely attributable to the indirect effect of UGPA on completion through graduate GPA. The total effects associated with GRE-P and GRE-Q are not statistically significant.}
\end{table}

Having demonstrated that graduate GPA (GGPA) is correlated with several quantitative admissions metrics of interest as well as final PhD completion, we now turn to the results of a mediation analysis to determine the extent to which GGPA mediates the relationship between these variables.  As described previously, these results yield insight into whether better performance in undergraduate coursework and on GRE examinations increase a student's likelihood of completing a PhD program directly or indirectly via GGPA.  Over the course of performing these analyses we explored numerous mediation models using different combinations of covariates (e.g. gender and race) to explore their effect on the results.  For instance, we probed the effects of including these variables as moderators in the analysis to account for varying predictive effects among different demographics.  However, results were consistent regardless of how these covariates were included in the model. 

We perform a separate mediation analysis for each predictor variable (UGPA, GRE-P, and GRE-Q).  For each analysis, we begin by calculating the direct and indirect effects of each metric on PhD completion using Eqs. (\ref{eq:indirecteffect}) and (\ref{eq:directeffect}).  We also calculate the total effect, which is the sum of the direct and indirect effects.  The total effect is comparable to the result one would obtain by using logistic regression to predict changes in probability of PhD completion as was done in \citeauthor{miller_typical_2019} \cite{miller_typical_2019}.

This procedure requires us to choose both a control and treatment value for the admissions metrics, since the output of the mediation analysis is the predicted difference in probability of PhD completion if a student who earned the control score had earned the treatment score instead.  

Table \ref{tab:exampleMediation} shows the results of a mediation analysis in which we calculate the predicted change in probability of PhD completion if a student had earned a 3.90 undergraduate GPA rather than a 3.46 undergraduate GPA.  This change corresponds to a shift from the 25th to 75th percentile of undergraduate GPAs in our data.  We observe that the predicted total effect of this change is to increase the student's overall probability of PhD completion by 6.0\% ($p<0.01$).  That effect is entirely attributable to the indirect effect of UGPA on PhD completion through graduate GPA, since the direct effect is estimated to be -0.1\% and is not statistically significant ($p=0.97$), while the indirect effect is estimated to be 6.0\% and is statistically significant ($p<0.01$).  Meanwhile, mediation analysis using GRE-P as the predictor variable estimates that due to a change from the 25th to 75th percentile of GRE-P scores among students in this study, the probability of PhD completion increases by 3.7\% (this change corresponds to a shift in GRE-P percentile ranking from 35 to 71 among the overall physics test-taking population).  This result just misses the threshold of statistical significance ($p=0.08$).  However, any effect that may be associated with a higher GRE-P score is attributable to the indirect effect of GRE-P on GGPA, which is estimated to be 4.2\% and is statistically significant ($p<0.01$).  For a change in GRE-Q score from the 25th to 75th percentile of scores among students in this study (a change in GRE-Q percentile ranking from 79 to 91 among the overall physics test-taking population), the direct (2.4\%, $p=0.13$) and indirect (0.9\%, $p=0.26$) predictive effects on PhD completion are not statistically significant.  Their sum, the total effect, estimates an increase in PhD completion probability of 3.4\%, and like the total effect of GRE-P also just misses the threshold for statistical significance ($p=0.07$).

\begin{figure*}
    \includegraphics[]{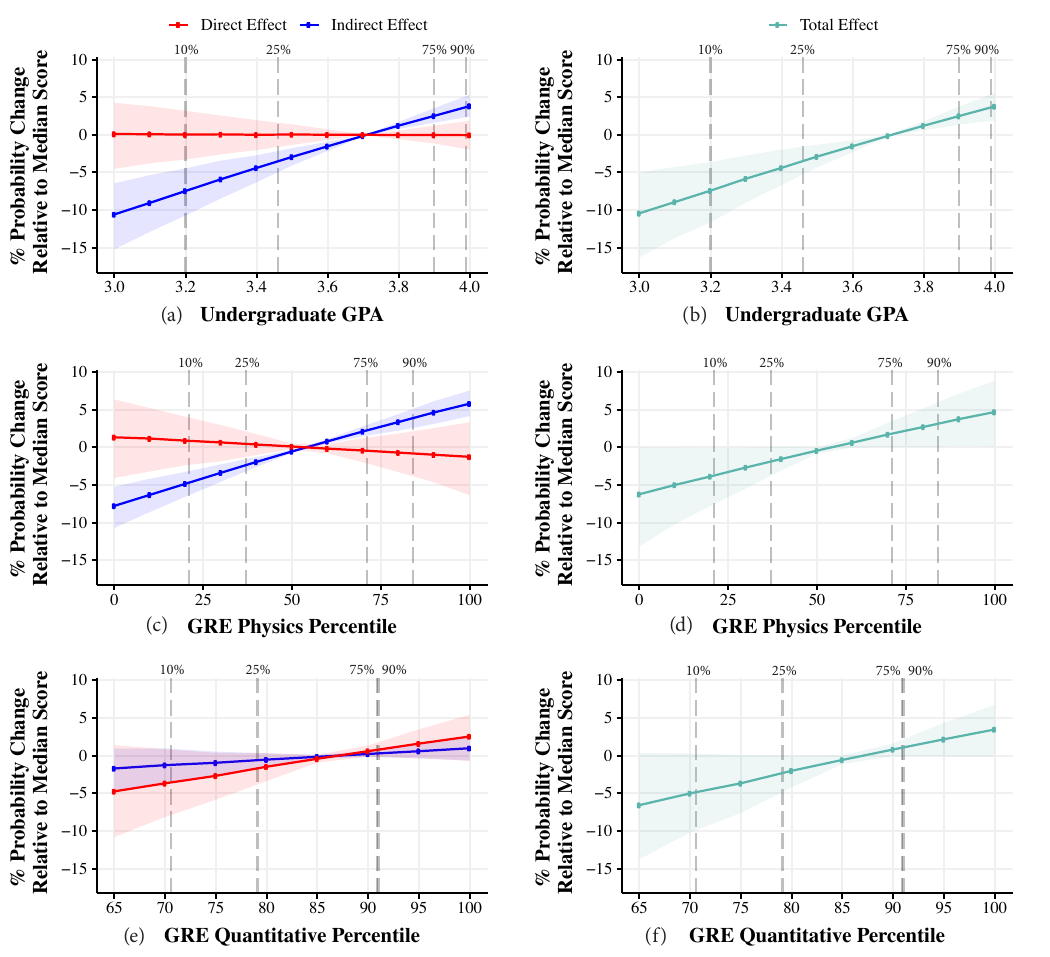}
    \caption{\label{fig:mediations} Mediation analysis results predicting a student's change in probability of completion, split into direct (red) and indirect via graduate GPA (blue) effects. Points on each plot represent individual calculations of probability change relative to each admission metric's median value.  95\% confidence intervals are indicated by the shaded region around each line; hence, if the shaded region contains the line $y = 0$,  the effect is not statistically significant at the $\alpha=.05$ level.  The median value of the $x$-axis variable is clearly shown on the plot as the point where the direct and indirect effect lines intersect, corresponding to a total probability change of 0\%.  The total effect of UGPA as well as the indirect effects associated with UGPA and GRE-P are statistically significant across all magnitudes of score change.}
\end{figure*}

To examine how the predicted probability of PhD completion changes over a broad range of UGPA and GRE scores, we repeat this single mediation analysis for a range of treatment values, as suggested in \cite{imai_general_2010}.  We opt to choose our data's median values as the baseline control value of each metric, then compute the direct and indirect effects for a variety of treatment values with respect to this baseline.  The result is a plot of predicted PhD completion probability change as each score is varied.  Figure \ref{fig:mediations} graphically summarizes the results of the mediation analysis.  Figures \ref{fig:mediations}a), c), and e) display the results of the three separate analyses predicting PhD completion probability changes relative to the median value of each independent variable.  Points on each plot represent individual calculations of probability change relative to each admission metric's median value.  Hence, the median value of the $x$-axis variable is clearly shown on the plot as the point where the direct (red) and indirect effect (blue) lines intersect, corresponding to a total probability change of 0\%. 

\begin{table*}[t]
\def\arraystretch{1.2}
\begin{tabularx}{\textwidth}{Y|YY|YY|}
\cline{2-5}
 & \multicolumn{2}{c|}{Current Study} & \multicolumn{2}{c|}{Previous Study (Ref. [1])} \\ \hline
\multicolumn{1}{|l|}{\begin{tabular}[c]{@{}l@{}}Independent\\ Variable\end{tabular}} & \begin{tabular}[c]{@{}c@{}}Total Effect\\ Value\\ (Probability Scale)\end{tabular} & \begin{tabular}[c]{@{}c@{}}Predicted Change in\\ Completion Probability\\ (25th-75th Percentile)\end{tabular} & \begin{tabular}[c]{@{}c@{}}Ref. [1]\\ Coefficient\\ (Log Odds Scale)\end{tabular} & \begin{tabular}[c]{@{}c@{}}Predicted Change in\\ Completion Probability\\ (25th-75th Percentile)\end{tabular} \\ \hline
\rowcolor[gray]{.9}[\tabcolsep] 
\multicolumn{1}{|l|}{\cellcolor[gray]{.9}UGPA} & 0.060 ($p<0.01$) & 6.0\% & 0.60 ($p<0.01$) & 3.6\% \\
\multicolumn{1}{|l|}{GRE-P} & 0.037 ($p=0.08$) & 3.7\% & $5 \times 10^{-3}$ ($p=0.09$) & 3.1\% \\
\rowcolor[gray]{.9}[\tabcolsep] 
\multicolumn{1}{|l|}{\cellcolor[gray]{.9}GRE-Q} & 0.034 ($p=0.07$) & 3.4\% & $10 \times 10^{-3}$ ($p=0.04$) & 2.1\% \\ \hline
\end{tabularx}
\caption{\label{tab:comparison} Comparison of results from the current study to results presented by \citeauthor{miller_typical_2019} \cite{miller_typical_2019} in which the authors predict PhD completion using a logistic regression model.  Despite using different statistical methods and omitting program tier as a predictor in our model, we observe that the results obtained in this study are qualitatively consistent with those reported previously. In both cases, UGPA is the strongest predictor of PhD completion among admissions metrics tested.}
\end{table*}

Percent probability changes due to direct effects of each admissions metric on PhD completion are shown in red, while indirect effects on PhD completion transmitted through GGPA are shown in blue.  Plots of the total effect of each variable on PhD completion probability, which is the sum of the direct and indirect effects, are shown in Figures \ref{fig:mediations}b), d), and f). The shaded ribbons around the lines representing the best-estimates of probability change show the 95\% confidence interval.  Hence, if this shaded region contains the line $y = 0$,  the effect is not statistically significant at the $\alpha=.05$ level.

In agreement with the results of the example analysis shown in Table \ref{tab:exampleMediation}, the total predictive effect of UGPA on PhD completion is statistically significant while the total effects of GRE-P and GRE-Q do not reach the threshold for statistical significance, indicated by the error bands that encompass the $y=0$ line (Figures \ref{fig:mediations}d and f).  Still, as suggested by the result in the Total Effect column of Table \ref{tab:exampleMediation} for GRE-P ($p=0.08$) and GRE-Q ($p=0.07$), these effects are close to reaching statistical significance; for reference, each point in Figure \ref{fig:mediations}d) the GRE-P confidence ribbon surpasses the line $y=0$ by less than 0.5\%.  Thus these results provide some evidence that scoring more highly on the GRE-P and GRE-Q are positively associated with higher rates of completion, although the effects are not statistically significant.   

Also consistent with the results of the example analysis shown in Table \ref{tab:exampleMediation} the results of the three mediation analyses indicate that predictive effects of UGPA and GRE-P on a student's PhD completion are entirely mediated by GGPA, defined by the fact that the indirect effect of these admissions metrics are statistically significant across all magnitudes of score change while their direct effects are not.  These indirect effects are shown in Figure \ref{fig:mediations} by the blue lines in the UGPA and GRE-P plots, whose error ribbons do not contain the line $y=0$.  The interpretation of this result is that UGPA effectively predicts a student's GGPA, which in turn predicts PhD completion.  Similarly, any increase in PhD completion probability associated with increases in GRE-P scores are a result of the indirect effect through graduate GPA, although the total effect is not statistically significant.    

With regard to GRE-Q, given the weak relationship between GRE-Q and GGPA revealed by the multiple regression analysis in Section \ref{subsec:RoleofGGPA}, it is unsurprising that the indirect effect shown in blue on Figure \ref{fig:mediations} is nearly zero.  Indeed, as indicated by the earlier results in Table \ref{tab:exampleMediation} any predictive effect from GRE-Q on completion appears to stem from the direct effect of GRE-Q on PhD completion, although the direct effect does not achieve statistical significance at the $\alpha=0.05$ level.
  
Lastly, despite using different statistical methods and omitting program tier as a predictor in our model, we observe that the results obtained in this study are qualitatively consistent with those reported by \citeauthor{miller_typical_2019} \cite{miller_typical_2019}.  Table \ref{tab:comparison} shows a comparison of the total effects predicted by the mediation model presented here with the results of the logistic regression model presented in \cite{miller_typical_2019}, again using as an example the predicted changes in probability of PhD completion due to shifts from the 25th to 75th percentile in score for the different admissions metrics.  In both analyses, the predictive effects associated with changes in UGPA are statistically significant and are the largest in magnitude among the tested metrics.  Effects associated with changes in GRE-P scores are not statistically significant at the $\alpha=0.05$ level in either model but are close ($p=0.08$ in this study and $p=0.09$ in \cite{miller_typical_2019}).  The only minor inconsistency between the two analyses is revealed in the results of the GRE-Q models.  Whereas in \cite{miller_typical_2019} the predictive effect of GRE-Q on completion was barely significant ($p=0.04$), it is not statistically significant here ($p=0.07$).  However, as demonstrated in Table \ref{tab:comparison}, effects associated with GRE-Q are not strong in either model and both are very close to the $\alpha=0.05$ threshold for statistical significance, indicating that the two results are still approximately consistent.



\section{\label{sec:Discussion}Discussion}
\subsection{\label{subsec:Implications} Interpretation of Results}

The results presented in Section \ref{subsec:MediationResults} give new insight into how admissions committees may contextualize the use of quantitative admissions metrics.  Clearly, no existing metric provides unassailable evidence that a student will complete their graduate program.  However, among the imperfect quantitative admissions metrics commonly used by admissions committees, the consistent message from this work and others is that undergraduate GPA offers the most promising insight into whether physics graduate students will earn a PhD.  Moreover, there is no significant difference between male and female applicants' UGPAs as there is for the GRE-P, meaning that its use in ranking applicants is less likely to skew diversity of admitted students.

As demonstrated by the multiple regression analyses in Section \ref{subsec:RoleofGGPA}, UGPA is most strongly associated with graduate course performance among the variables tested.  In some ways this is an expected result: UGPA, the metric that directly measures a student's in-class performance, is most effective at predicting future in-class performance in graduate school.  Still, UGPA would seem to vary greatly depending on the student's particular undergraduate institution while a standardized exam like the GRE-P is consistent across all students.  It is possible that UGPA may also be signaling socio-emotional skills such as achievement orientation and conscientiousness, which are known to predict high levels of performance both in and out of the classroom \cite{schmidt1998validity, shultz2012admission, lievens2012validity, victoroff2013relationship}.  These sorts of socio-emotional skills are also shown in the relevant research literature to lack the race, gender, and culture of origin gaps that are found on many standardized tests \cite{oswald2011personality, feingold1994gender, emmerling2012emotional}.  

While previous work showed that UGPA was an effective predictor of PhD completion, mediation analysis demonstrates that relationship is entirely transmitted through UGPA's ability to predict GGPA.  Thus, although UGPA is the best predictor of a student's final disposition, our analysis indicates that it is not a direct measure of PhD completion.  Rather, the observed relationship between undergraduate grades and completion is explained by the intervening variable GGPA.  Regarding the magnitude of the observed effects, we see that changes in score from the 25th to 75th percentile in UGPA (3.46 to 3.90) are associated with an 6\% increase in completion probability.  Changes across a broader range of  UGPAs from the 10th to 90th percentile (3.2 to 3.98), the mediation model predicts an 11\% increase in PhD completion probability.

Multiple regression and mediation analyses also yield improved insight into the information provided by GRE-P.  Consistent with prior published work by ETS \cite{kuncel2001comprehensive, gre1989validity}, regression analysis reveals that GRE-P is an effective predictor of graduate grades.  However, the effect associated with UGPA is approximately 50\% larger than the effect associated with GRE-P.  Regarding the relationship between GRE-P scores and PhD completion, any association between GRE-P performance and PhD completion is entirely mediated by graduate course performance, similar to UGPA.  This particular indirect effect indicates that a student who scores more highly on the GRE-P is more likely to perform better in graduate school courses, which may slightly improve their probability of graduation (although the total effect of GRE-P is not statistically significant).  However, this indirect effect is still smaller than the indirect effect  associated with UGPA, as illustrated in Figure \ref{fig:mediations}.  

Notably, despite the existence of a large gender gap in GRE-P scores (the median GRE-P percentile for females is 35 and 57 for males), male and female graduate students earn nearly indistinguishable graduate grades (Figure \ref{fig:boxplots}).  Moreover, there is no practical relationship between gender and PhD completion (Table \ref{tab:correlations}).  We have also done a preliminary analysis of this same data examining the relation between gender and time to PhD completion, and found that no statistically significant difference exists in the time it takes for male and female physics graduate students to complete doctoral degrees.  The disparity in GRE-P scores between male and female test takers is therefore anomalous, as it does not appear to be related to differences in ability or level of preparation and is not reflected in subsequent graduate performance.

Results of multiple regression and mediation analyses that show GRE-Q is not strongly associated with increased graduate course performance are unsurprising given the GRE-Q's task relevance is lower than subject tests and undergraduate grades.  Indeed, in its Guide to the Use of Scores\cite{ets2020guide}, ETS describes the GRE-Q as testing ``high school mathematics and statistics at a level that is generally no higher than a second course in algebra; it does not include trigonometry, calculus or other higher-level mathematics."  As shown in Table \ref{tab:exampleMediation}, the association between GRE-Q scores and PhD completion just misses the $\alpha=0.05$ threshold of statistical significance ($p=0.07$).  Combined with previous studies in which this weak association was statistically significant \cite{miller_typical_2019}, evidence suggests a weak relationship between GRE-Q scores and completion.  

Both direct and indirect effects of GRE-Q on completion were not statistically significant and therefore we cannot discern with certainty which small effect is more important.  Considering the case of a possible direct effect between GRE-Q and completion, the low task relevance makes it unlikely that GRE-Q is a measure of research competence or perseverance.  One hypothesis is that socioeconomic status (SES) could be confounding the relationship between GRE-Q, an exam consisting of high school level mathematics questions, and PhD completion.  Students with lower SES may have fewer opportunities academically and may perform worse on a standardized exam like the GRE-Q.  Indeed, it is estimated that roughly 20\% of variance in standardized test scores can be explained by SES \cite{sackett_role_2012}  In general, previous research \cite{owens2020} indicates that SES impacts whether students possess the proper resources to support them should financial, health or other external circumstances make it difficult to complete.  Moreover, doctoral students from lower social classes are more likely to experience a lower sense of belonging in graduate school, often due to the residual financial burdens that are not mitigated by graduate stipends \cite{ostrove2011social}.  Reduced sense of belonging in graduate school drives lower interest in pursuing advanced careers in the field, and ultimately a lower likelihood of completing a PhD.



\subsection{\label{subsec:Limitations} Limitations and future research}
\subsubsection{Assumptions in Causal Mediation Analysis}

The causal effect framework laid out in Section \ref{subsec:Mediation} requires several assumptions to be satisfied in order for effect estimates to be properly identified.  All of the assumptions underlying causal mediation analysis refer to ``confounding variables," which are variables that influence two other variables simultaneously, thereby causing a spurious association between them.  In this section we discuss whether it is plausible that our study has satisfied these assumptions, as well as suggestions for future researchers seeking to perform similar analyses.

\begin{figure}[t]
    \includegraphics[]{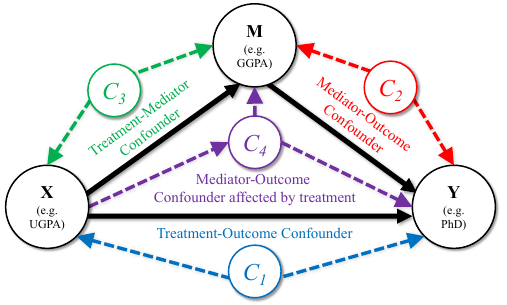}
    \caption{\label{fig:assumptions} Mediation diagram showing the relationship of ``confounding variables" $C_{1\text{-}4}$ to the variables of interest $X$, $Y$, and $M$.  To satisfy the underlying assumptions of causal mediation analysis, researchers must account for these confounding variables in their regression models.}
\end{figure}

Although the assumptions prescribed by causal mediation analysis may be stated in multiple ways \cite{imai_identification_2013, pearl2014interpretation}, we describe them in the manner presented by \cite{vanderweele_explanation_2015}, who condenses them into four requirements.  These assumptions, displayed graphically in Figure \ref{fig:assumptions}, are that there exists (1) no unmeasured treatment-outcome confounders, (2) no unmeasured mediator-outcome confounders, (3) no unmeasured treatment-mediator confounders, and (4) no mediator-outcome confounder affected by the treatment.

The final assumption is equivalent to assuming that there is not an alternative mediating variable for which we have not accounted \cite{imai_identification_2013}.  Concerns about this assumption are handled using methods for multiple mediators, discussed in detail in the Supplemental Material.  

Assumptions 1-3 require us to be sure we have included all variables that could be confounding the relationships between $X$ (UGPA, GRE-P or GRE-Q), $Y$ (PhD completion), and $M$ (GGPA) in our regression models. These are shown graphically by variables $C_1$, $C_2$, and $C_3$ in Figure \ref{fig:assumptions}.  Some of the $C$ variables may influence more than one of $X$, $Y$, and $M$, or may influence each other, so categorization into $C_1$, $C_2$, and $C_3$ is not exclusive.  Accounting for variables $C_1$ and $C_3$ correspond to assumptions normally made in observational studies to calculate total effects \cite{vanderweele_mediation_2016}, while accounting for variables labeled $C_2$ is important specifically for the estimation of direct and indirect effects. 

\subsubsection{Sensitivity Analysis for Violation of Assumptions}
Unfortunately, as is the case in any regression analysis, whether these assumptions are met is not testable; however, there exist several ways to probe the robustness of our findings under certain violations.  In particular, sensitivity analyses have been developed \cite{imai_general_2010, vanderweele_explanation_2015} to determine how strong a confounding effect between the mediator and the outcome would have to be in order to make statistically significant effects become no longer significant.

We perform such sensitivity analyses on each of the imputed data sets used in the single mediator models presented in Section \ref{subsec:MediationResults} to assess their robustness.  We then report the results of the sensitivity analyses averaged across the imputed datasets.    

Let $R^2_M$ and $R^2_Y$ represent the proportions of original variances explained by the unobserved confounder for the mediator GGPA and the outcome PhD completion, respectively.  The result of a sensitivity analysis is a single value representing the product $R^2_M \times R^2_Y$ that identifies the amount of original variance in the mediator and outcome that the confounder would have to explain in order to make the observed effect vanish. Hence, the sensitivity analysis results in a family of solutions for which the equation $R^2_M \times R^2_Y = constant$ is satisfied.

Our sensitivity analysis reveals that for a confounder to explain enough variance to make the indirect effect of UGPA on GGPA vanish, the product $R^2_M \times R^2_Y$ would have to be, on average, 0.072 (standard deviation = 0.006).  For context, the variables included in the mediation models presented here, as a group, are able to explain 11\% of variance in the GGPA outcome ($R^2 = 0.11$); for PhD completion, $R^2 = 0.32$.  

Comparing these values to those required to satisfy the equation $R^2_M \times R^2_Y = 0.072$, we see that a mediator-outcome confounder would have to be a better predictor of both GGPA and PhD completion than any quantity included in the model so far.  For instance, suppose a mediator-outcome confounder satisfied the equation $R^2_M \times R^2_Y = 0.15 \times 0.45 = 0.072$, sufficient to nullify the indirect effect of UGPA on completion. The relationship between the confounder and GGPA ($R^2_M = 0.15$) would be larger than all other variables combined in the model.  The link between the confounder and PhD completion ($R^2_Y=0.45 $) would be larger than all other variables combined in the model as well. 

Since the values of $R^2_M$ and $R^2_Y$ must be so high relative to the rest of the variables in our model to nullify the indirect effect of UGPA on PhD completion, this result appears to be fairly robust.

Results of sensitivity analyses on the indirect effect of GRE-P on PhD completion were similarly robust: for a confounder to explain enough variance to make the indirect of GRE-P on GGPA vanish, the product $R^2_M \times R^2_Y$ on average would have to be 0.069 (standard deviation = 0.001).  This is essentially the same as the analyses on the robustness of the UGPA result, so its interpretation is the same as above.
 
More details regarding the sensitivity analyses performed, including contour plots of $R^2_M \times R^2_Y$ products for which indirect effects of UGPA and GRE-P vanish, are available in the Supplemental Material.

\subsubsection{Conceptualizing possible unmeasured confounders}

Thoroughly considering variables that could conceivably act as confounders in Figure \ref{fig:assumptions} would benefit future researchers studying graduate admissions.  Non-quantitative aspects of a student's admission credentials such as letters of recommendation and prior research experience may be associated with PhD completion, and could represent several $C$ variables in the diagram.  For example, prior research experience could act as a mediator-outcome confounder, labeled $C_2$ in Figure \ref{fig:assumptions}.  A student with more research experiences prior to entering graduate school may be more likely to complete a PhD than a student with fewer research experiences, since they have already become familiar with the expectations associated with scientific research and have already demonstrated the motivation to pursue it independently.  More research experience may also translate to better graduate course performance, particularly if graduate courses are well-aligned with the goal of preparing students for future research.

As discussed earlier, information on student socioeconomic status could be useful as well.  We also see SES as potentially representing several confounding relationships.  For instance, it could act as a treatment-outcome confounder, influencing both GRE scores and PhD completion.  It is estimated that roughly 20\% of variance in standardized test scores can be explained by SES \cite{sackett_role_2012}, so SES could be influencing performance on tests such as the GRE-Q and GRE-P. Students with lower SES may have fewer resources to support them should financial, external circumstances arise that make it difficult for them to complete their PhD \cite{owens2020, owens2018}.  Including data on these and other possible confounders would bolster causal claims in future analyses.  Furthermore, our current models only explain a small amount of the overall variance in the outcome variable, and many other factors are surely at play.

Lastly, although graduate course performance and PhD completion represent some aspects of graduate school success (as evidenced by their inclusion in GRE validation studies), they are certainly crude metrics.  Future work should explore other outcomes as well, including success measures such as research productivity, job attainment, or graduate student satisfaction.

\section{\label{sec:Conclusion}Conclusions}

Using data visualization, regression analyses, and mediation analyses, we investigated the role that graduate GPA plays on a physics graduate student's path to PhD completion.  We aimed to answer two primary research questions: 1) How do commonly used admissions metrics and demographic factors relate to physics graduate GPA?, and 2) What role does graduate GPA play in predicting PhD completion, and does it mediate the influence of these other predictor variables on PhD completion?  Broadly, we find that across the dynamic range of scores in the data, undergraduate GPA was a better predictor of both graduate GPA and final disposition than GRE scores.  

Regarding the first research question of how various admissions metrics and demographic factors relate to physics graduate GPA, we see that significant but weak predictive effects at the 95\% threshold were found for the numerical metrics undergraduate GPA ($\beta = 0.24$, $t=8.88$, $p<.01$) and GRE Physics ($\beta = 0.15$, $t=5.02$, $p<.01$); GRE Quantitative and Verbal scores are not significantly associated with graduate GPA.  The regression model predicts that for a 0.10 score increase in undergraduate GPA, a student's graduate GPA is expected to increase on average by 0.035 points, holding all other predictors fixed.  Meanwhile, a 10 percentile increase in GRE Physics score is associated with a 0.031 point increase in graduate GPA on average, again holding other predictors fixed.  For comparison, a change in UGPA from the 25th to 75th percentile of scores in our data predicts a 0.15 point increase in graduate GPA, whereas a change in GRE-P from the 25th to 75th percentile of scores in our data predicts a 0.11 increase in graduate GPA.    

We also observe that the graduate GPAs are not statistically different between males and females. Hence, the statistically significant gap in performance by gender on the GRE Physics exam (within our data the median GRE Physics percentile for females is 35 and 57 for males) does not carry over to subsequent graduate course performance.  The large difference in performance is unexplained, yet is potentially problematic for promoting diversity in physics graduate school.  Multiple regression analysis did not reveal race to be a statistically significant predictor of graduate GPA, but unfortunately it is difficult to properly interpret relationships between race and graduate grades due to a small $N$.  Small sample size precludes useful interpretation of the results pertaining to Black, Hispanic, Native American, and Asian students. 

As to the second research question of whether graduate GPA mediates the influence of these other predictor variables on PhD completion, we find that UGPA predicts PhD completion indirectly through graduate grades.  Only UGPA is a statistically significant predictor of overall PhD completion (a change in UGPA from the 25th to 75th percentile of scores in our data predicts a 6\% increase in PhD completion probability, $p<0.01$), and that effect is entirely attributable to the indirect effect of UGPA on PhD completion through graduate GPA.  The indirect effect associated with UGPA on PhD completion was statistically significant across all magnitudes of score changes, while the direct effect was not (see Figure \ref{fig:mediations}). Thus UGPA effectively predicts graduate course performance, which is then associated with degree completion.  The association between GRE-P scores and PhD completion is not statistically significant (a change in GRE-P from the 25th to 75th percentile of scores in our data predicts a 3.7\% increase in PhD completion probability, $p=0.08$).  However, like UGPA, the indirect effect associated with increases in GRE-P score was also statistically significant across all magnitudes of score change, meaning that any predictive effect that GRE-P score may have is therefore also linked indirectly through graduate GPA.

Although these models explain some of the variance in student outcomes (the variables included in the mediation models, as a group, explain 11\% of variance in graduate GPA ($R^2 = 0.11$); for PhD completion, $R^2 = 0.32$), much of the variation lies in factors outside the models, in both unmeasured student characteristics prior to admission and unmeasured aspects of the graduate student experience. 

No standardized test measures the research and project management skills it takes to successfully complete a multi-year research project, yet those are the skills that are so highly valued in PhD graduates.  The GRE-P utilizes two-minute theoretical physics problems to ascertain aspects of students’ physics knowledge, but it neglects the broad range of computational and experimental skills used in contemporary physics research.  Because undergraduate GPA reflects a mix of courses that include theory, experiment, computation, and in some cases research projects, it could be a more useful measure of research-relevant skills. However, our result that UGPA only indirectly predicts PhD completion seems to indicate those research-relevant skills are not a major part of the overall UGPA.  Identifying a broader set of applicant characteristics that predict graduate student outcomes is essential.

By better understanding and improving graduate education, we have the opportunity to meet societal goals of a highly-skilled advanced STEM workforce that reflects the diversity of our society. While adjusting admissions practices may offer some improvements by adjusting which students are allowed to undertake graduate study, such efforts do nothing to improve the graduate student experience and train graduate students more effectively for STEM careers. The potential for innovation and improvement within graduate education is large and is the area deserving substantial increased attention for education research and programmatic implementation. While our study gives admissions committees greater insight into how and why various quantitative scores link to completion, our discussion of the limitations also points to areas where future researchers can build.  We encourage the continued study of not only the physics graduate admissions process, but also the ongoing experience of students in PhD programs, how they are taught, mentored, and supported through their growth as individuals within a larger scientific community.

\begin{acknowledgements}
The authors are pleased to acknowledge valuable discussions with Nicholas Young, Julie Posselt, and Rachel Silvestrini.  This work was supported by NSF grants 1633275 and 1834516. 
\end{acknowledgements}


\bibliography{ms.bib}

\end{document}


\preprint{}

\title{Supplementary Materials:\\Analyzing admissions metrics as predictors of graduate GPA and whether graduate GPA mediates PhD completion}
\author{Mike Verostek$^{(1,2)}$, Casey Miller$^{(2)}$ and Benjamin Zwickl$^{(2)}$}
\affiliation{$^{(1)}$Department of Physics and Astronomy, University of Rochester}

\affiliation{$^{(2)}$Department of Physics and Astronomy, Rochester Institute of Technology}

\maketitle

\section{Multiple mediator model results}

\begin{figure}[H]
    \centering
    \includegraphics[]{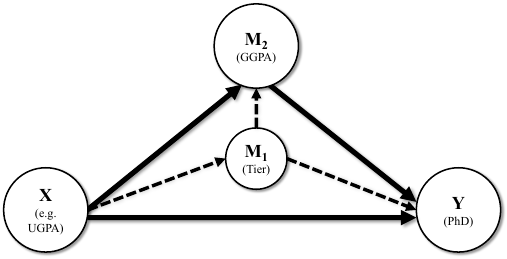}
    \caption{\label{fig:multiple_mediators} Qualitative diagram displaying the proposed effects in a multiple mediator model that includes program tier.}
\end{figure}

The results presented in the main analysis (see Section IV B) utilize models that suppose the existence of one mediating variable (GGPA).  Single mediator models such as this have been studied extensively \cite{vanderweele_mediation_2014}, and provide a well-established mathematical framework (see Section III C of the main text) to calculate direct and indirect effects.  However, it is sometimes of interest to consider cases involving multiple mediators.

In Section IV A of the main text we found that students attending institutions categorized as Tier 3 earned GGPA scores that were statistically lower than students attending Tier 1 or 2 institutions.  We suppose that program tier is a proxy measure for institutional resources and support, meaning that a higher tier institution might possess greater academic resources, which could improve student graduate performance.  Better resources for students could also directly affect PhD completion, as indicated in \cite{miller_typical_2019}.  Hence, we also consider a multiple mediator model corresponding to Figure \ref{fig:multiple_mediators} in which program tier is posited to be a second mediating variable between admissions metrics and PhD completion.

Methods to disentangle the effect that each individual path in a multiple mediator model has on the dependent variable has been the subject of much recent research \cite{daniel_causal_2015, imai_identification_2013}.  We follow the method provided in \cite{steen_flexible_2017}, which extends the counterfactual mediation analysis framework to deal with particular cases of multiple mediators. 

\begin{figure*}
    \includegraphics[]{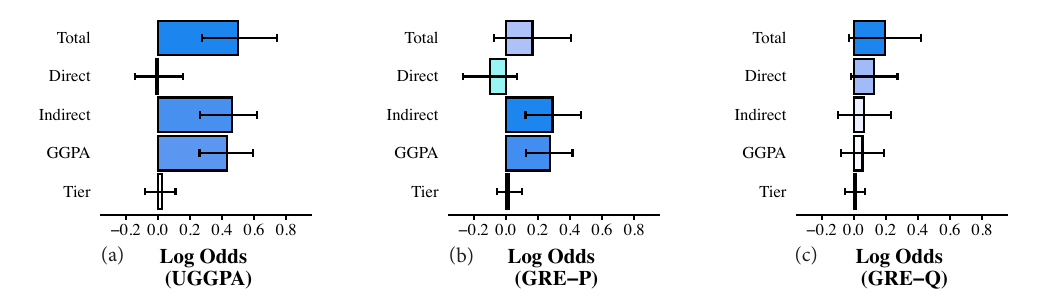}
    \caption{\label{fig:joint_mediations} Results of a model using multiple mediators to predict PhD completion.  Bars represent the change in log odds associated with a one standard deviation change in the independent variable.  Results are qualitatively the same as the single mediator case, with graduate GPA. Despite its inclusion, the indirect effects transmitted through program tier are not statistically significant in any of the multiple mediator models and are minuscule compared to the effects associated with GGPA.}
\end{figure*}

The results of this analysis are direct and indirect effects of each admissions metric on PhD completion, with the indirect effect broken down into the portion transmitted through GGPA and the portion transmitted through program tier.  Indeed, the output is qualitatively similar to the single mediator case described previously: direct and indirect effects are calculated, and effects for which the 95\% error bars do not cross zero are statistically significant.  These results are summarized in Figure \ref{fig:joint_mediations}.  

The single mediator results presented in Section IV B of the main text are fully consistent qualitatively with the results of this multiple mediator analysis. Again, only UGPA has a significant total effect on PhD completion, and that effect is fully transmitted through GGPA.  The indirect effect of GRE-P is significant, meaning that increased GRE-P scores tend to predict better GGPA scores, which in turn predict higher PhD completion.  Still, this effect lesser in magnitude than the effect of UGPA.  Again, none of the predictive effects of GRE-Q are significant, although the direct predictive effect of GRE-Q on completion borders closely on statistical significance.  

Quantitatively the results of multiple mediation analysis yield effect estimates on the log odds scale rather than a probability scale.  Exponentiation of the log odds converts the scale to a regular odds ratio.  Thus, the numerical estimates for the effects shown in Figure \ref{fig:joint_mediations} may be interpreted as the change in the log odds ratio (or if exponentiated, the odds) of PhD completion due to a one standard deviation increase in the independent variable.  For instance, the ``Indirect" bar of Figure \ref{fig:joint_mediations}a  shows the best estimate for the log odds ratio for the indirect predictive effect of UGPA on PhD completion is 0.46 (meaning that the odds ratio is exp(0.46)=1.58).  Thus, the model predicts that a student who had scored one standard deviation higher in their UGPA would increase their odds of PhD completion by 58\%.

Although this may seem like a large value, recall that the probability of an event is given by $odds/(1+odds)$, meaning that the larger the probability, the larger the difference in odds.  Models in \cite{miller_typical_2019} estimated that probability of PhD completion was typically around 75\% for most U.S. students, so an increase of approximately 50\% in odds of completion would result in a probability increase of about 7\%.  This estimate falls within the 95\% confidence interval of the results presented in the single mediator model for a one standard deviation increase in UGPA  (approximately an increase in score of 0.3).  

Despite its inclusion, the indirect effects transmitted through program tier are not statistically significant in any of the multiple mediator models and are minuscule compared to the effects associated with GGPA.  For comparison, in the model using UGPA as the independent variable the increase in the log odds ratio of PhD completion associated with Tier is 0.03, while the log odds ratio increase associated with GGPA is 0.43.  Thus, the results of all multiple mediator models analyzed indicate that tier is a far weaker mediator than GGPA.  

\section{Multiple Regression and Multicollinearity}

To address potential concerns with multicollinearity in our primary regression analysis, we present here the results of several alternative regression models.  Multicollinearity is a phenomenon in which two or more predictors in a regression model are highly correlated \cite{daoud_multicollinearity_2017}.  When multicollinearity is present, the standard errors of the affected variables can become inflated \cite{mcclendon_multiple_2002, farrar1967multicollinearity}.  Increased standard errors may cause some variables to appear statistically insignificant when they should be significant.

Diagnosing multicollinearity often consists of examining bivariate correlations between variables in the regression model and calculating variance inflation factor (VIF) scores.  VIF scores are measures of the extent to which the variance of a regression coefficient is increased due to collinearity.  When two predictors are independent (orthogonal, zero correlation) their VIF score is 1.  As the correlation between the variables increase, so does the VIF score.

Unfortunately the threshold values of metrics for identifying multicollinearity are ambiguous \cite{mela2002coll}.  For instance, some authors suggest that correlations above $r=0.35$ \cite{tull_marketing_1990} are cause for concern, while some use $r=0.5$ as a cutoff \cite{donath2012predictors}. Others assert that $r=0.80$ or $r=0.90$ are likely to be problematic \cite{judge_introduction_1988, field_discovering_2012}.  Reviews of articles discussing multicollinearity thresholds indicate that it is most common for authors to use $r=0.70$ to $r=0.80$ and VIF $>10$ rule-of-thumb cutoffs for investigating possible multicollinearity \cite{dormann_collinearity_2013, vatcheva_multicollinearity_2016}.

Effects due to multicollinearity and variance inflation are negligible in our analyses.  As noted in the main text, we considered the possibility that it would be inappropriate to include both GRE-P and GRE-Q in the same model due to collinearity concerns.  However, both the correlation coefficient and VIF scores for these variables fail to meet the most commonly used thresholds.  The bivariate correlation ($r_{xy}=0.51$) between the two is not high enough to warrant genuine concern.  Furthermore, the variance inflation factor (VIF) of both GRE-P and GRE-Q was below 1.75 in each imputed dataset's regression model, well below the commonly cited threshold of 10.

\begin{table}[]
\def\arraystretch{1.2}
\begin{tabular}{p{1.5cm}|>{\centering\arraybackslash}p{1cm}|}
\cline{2-2}
                             & VIF  \\ \hline
\multicolumn{1}{|c|}{UGPA}   & 1.16 \\ \hline
\multicolumn{1}{|c|}{GRE-P}  & 1.56 \\ \hline
\multicolumn{1}{|c|}{GRE-Q}  & 1.47 \\ \hline
\multicolumn{1}{|c|}{GRE-V}  & 1.18 \\ \hline
\multicolumn{1}{|c|}{Race}   & 1.05 \\ \hline
\multicolumn{1}{|c|}{Gender} & 1.01 \\ \hline
\end{tabular}
\caption{\label{tab:VIFscores} Average Variance Inflation Factors (VIF) for each predictor in a multiple regression predicting graduate GPA across 20 imputed data sets.}
\end{table}

Still, we perform several analyses in order to explore the effect of eliminating either GRE-Q or GRE-P from our analyses.  We also consider a model in which we generate a composite GRE-Average score, which is the simple average of the two exam scores.  These are shown in Table \ref{tab:SeveralRegressions}.  The results of the main text's analysis are shown in Table \ref{tab:regressionMAINTEXT} for comparison.

Notably, as suggested by the low VIF scores, the standard errors of the GRE-Q and GRE-P coefficients remain approximately constant regardless of whether one or both variables are included in the regression model.  In the model presented in the primary analysis shown in Table \ref{tab:regressionMAINTEXT}, the standard errors for GRE-P and GRE-Q are $6 \times 10^{-4}$ and $12 \times 10^{-4}$, respectively.  When GRE-P is removed from the model, the standard error of GRE-Q decreases slightly to $11 \times 10^{-4}$.  When GRE-Q is removed, the standard error of GRE-P remains $6 \times 10^{-4}$. 

Across all versions of the multiple regression analyses, UGPA is highly significant and is the most effective predictor of GGPA among tested variables, evidenced by examining the standardized regression coefficients in each model.  Indeed, even in the case that GRE-P and GRE-Q are averaged into one score to consolidate their predictive ability we observe that UGPA is the most effective predictor.  This again reinforces the conclusion that undergraduate grades are the best predictors of graduate school course performance.

In the model with GRE-Q removed, the GRE-P coefficient is 0.17, which hardly changes from the model in which both GRE-Q and GRE-P are included (0.15).  Thus the practical effect on the GRE-P coefficient is negligible regardless of whether the model includes GRE-Q.  Meanwhile, exclusion of the GRE-P variable does indeed cause the GRE-Q coefficient become a statistically significant predictor.  This is unsurprising given the fact that its magnitude dropped from 0.09 to 0.03 with the inclusion of GRE-P while maintaining approximately the same standard error.        

\begin{table}[h]
\centering
\def\arraystretch{.8}
\begin{tabular}{llcc}
\multicolumn{4}{c}{Multiple Regression Results (*$p<.05$, **$p<.01$)}                                                                       \\ \hline\hline
\multicolumn{2}{l}{}                            & \multicolumn{2}{c}{Dependent Variable - GGPA} \\ \cline{3-4} 
\multicolumn{2}{l}{\begin{tabular}[c]{@{}l@{}}Independent\\ Variable\end{tabular}} &
  \begin{tabular}[c]{@{}c@{}}Coefficient\\ (Standard Error)\end{tabular} &
  \begin{tabular}[c]{@{}c@{}}Standardized\\ Coefficient\end{tabular} \\ \hline

\rowcolor[gray]{.9}[\tabcolsep]   
\multicolumn{2}{l}{Intercept} & 1.92** (0.15) & -0.06 \\

\rowcolor[gray]{.9}[\tabcolsep] 
\multicolumn{2}{l}{UGPA} & 0.35** (0.04) & 0.24** \\

\rowcolor[gray]{.9}[\tabcolsep] 
\multicolumn{2}{l}{GRE-P} & $31 \times 10^{-4}$** ($6 \times 10^{-4}$) & 0.15**          \\

\rowcolor[gray]{.9}[\tabcolsep] 
\multicolumn{2}{l}{GRE-Q} & $16 \times 10^{-4}$ ($12 \times 10^{-4}$) & 0.03              \\

\rowcolor[gray]{.9}[\tabcolsep] 
\multicolumn{2}{l}{GRE-V} & $3 \times 10^{-4}$ ($6 \times 10^{-4}$) & 0.01              \\

\multicolumn{2}{l}{Black} & -0.11 (0.09) & -0.23              \\

\multicolumn{2}{l}{Hispanic} & -0.01 (0.07) & -0.02  \\

\multicolumn{2}{l}{Native Am.} & 0.01 (0.23) & 0.03  \\

\multicolumn{2}{l}{Asian}  & -0.02 (0.06)  & -0.05  \\

\multicolumn{2}{l}{Other} & -0.24* (0.11)  & -0.51*  \\

\multicolumn{2}{l}{Undisclosed} & 0.07** (0.02)  & 0.15**  \\

\multicolumn{2}{l}{Gender} & 0.06* (0.03)  & 0.13* \\ \hline\hline

\multicolumn{2}{l}{\textit{N}} & 1955 &          \\

\multicolumn{2}{l}{Adjusted \textit{R}-Squared} & 0.11                      &             
\end{tabular}
\caption{\label{tab:regressionMAINTEXT} Coefficients of a multiple regression analysis modeling graduate GPA as a function of common quantitative admissions metrics.  Reference categories are White for race and Male for gender. These results are presented in the main analysis and are included here for comparison with alternate regression models.}
\end{table}

\begin{table*}[h]
\centering
\resizebox{\textwidth}{!}{%
\begin{tabular}{llcccccc}
\multicolumn{8}{c}{Multiple Regression Results (*$p<.05$, **$p<.01$)} \\ 
\hline\hline
\multicolumn{2}{l}{}                            & \multicolumn{6}{c}{Dependent Variables}                                                 \\ \cline{3-8} 
\multicolumn{2}{l}{}                            & \multicolumn{2}{c}{Model w/ No GRE-P} & \multicolumn{2}{c}{Model w/ No GRE-Q} & \multicolumn{2}{c}{Model w/ Composite GRE Score} \\ \cline{3-8} 

\multicolumn{2}{l}{\begin{tabular}[c]{@{}l@{}}Independent\\ Variable\end{tabular}} &
  \begin{tabular}[c]{@{}c@{}}Coefficient\\ (Standard Error)\end{tabular} &
  \begin{tabular}[c]{@{}c@{}}Standardized\\ Coefficient\end{tabular} &
  \begin{tabular}[c]{@{}c@{}}Coefficient\\ (Standard Error)\end{tabular} &
  \begin{tabular}[c]{@{}c@{}}Standardized\\ Coefficient\end{tabular} &
  \begin{tabular}[c]{@{}c@{}}Coefficient\\ (Standard Error)\end{tabular} &
  \begin{tabular}[c]{@{}c@{}}Standardized\\ Coefficient\end{tabular} \\ \hline

\multicolumn{2}{l}{Intercept} & 1.72** (0.15) & -0.03 & 2.00** (0.15) & -0.06 & 1.86** (0.15) & -0.06 \\

\multicolumn{2}{l}{UGPA} & 0.39** (0.04) & 0.26** & 0.36** (0.04) & 0.24** & 0.35** (0.04) & 0.23** \\

\multicolumn{2}{l}{GRE-V} & $7 \times 10^{-4}$ ($6 \times 10^{-4}$) & 0.03 & $4 \times 10^{-4}$ ($6 \times 10^{-4}$) & 0.02 & $2 \times 10^{-4}$ ($6 \times 10^{-4}$) & 0.01 \\

\multicolumn{2}{l}{Black} & -0.13 (0.09) & -0.26 & -0.11 (0.09) & -0.22 & -0.10 (0.09) & -0.21 \\

\multicolumn{2}{l}{Hispanic} & -0.01 (0.07) & -0.03 & -0.01 (0.07) & -0.03 & $-26 \times 10^{-4}$ (0.07) & -0.01 \\

\multicolumn{2}{l}{Native Am.} & -0.01 (0.24) & -0.02 & 0.02 (0.24) & 0.04 & $38 \times 10^{-4}$ (0.24) & 0.01 \\

\multicolumn{2}{l}{Asian} & $12 \times 10^{-4}$ (0.06) & $26 \times 10^{-4}$ & -0.02 (0.06) & -0.04 & -0.02 (0.11) & -0.04 \\

\multicolumn{2}{l}{Other} & -0.23* (0.11) & -0.49* & -0.25* (0.11) & -0.51* & -0.24* (0.11) & -0.50* \\

\multicolumn{2}{l}{Undisclosed} & 0.06* (0.02) & 0.12* & 0.07* (0.02) & 0.15* & 0.07* (0.02) & 0.15* \\

\multicolumn{2}{l}{Gender} & 0.01 (0.03) & 0.03 & 0.06* (0.03) & 0.13* & 0.06* (0.03) & 0.12* \\

\rowcolor[gray]{.9}[\tabcolsep]
\multicolumn{2}{l}{GRE-Q} & $43 \times 10^{-4}$** ($11 \times 10^{-4}$) & 0.09** & -- & -- & -- & -- \\

\multicolumn{2}{l}{GRE-P} & -- & -- & $34 \times 10^{-4}$** ($6 \times 10^{-4}$) & 0.17** & -- & -- \\

\rowcolor[gray]{.9}[\tabcolsep]
\multicolumn{2}{l}{GRE-Average} & -- & -- & -- & -- & $55 \times 10^{-4}$** ($9 \times 10^{-4}$) & 0.17** \\

\hline\hline

\multicolumn{2}{l}{\textit{N}} & 1955 & & 1955 & & 1955 & \\

\multicolumn{2}{l}{Adjusted \textit{R}-Squared} & 0.09 & & 0.11 & & 0.11 & \\

\end{tabular}}
\caption{\label{tab:SeveralRegressions} Coefficients of several multiple regression analysis modeling graduate GPA using different combinations of GRE scores.  Across all versions of the multiple regression analyses, UGPA is highly significant and is the most effective predictor of GGPA, evidenced by examining the standardized regression coefficients in each model  Reference categories are White for race and Male for gender.}
\end{table*}

\clearpage
\section{Mediation with Categorical Predictors}

\begin{figure*}[h]
\includegraphics[]{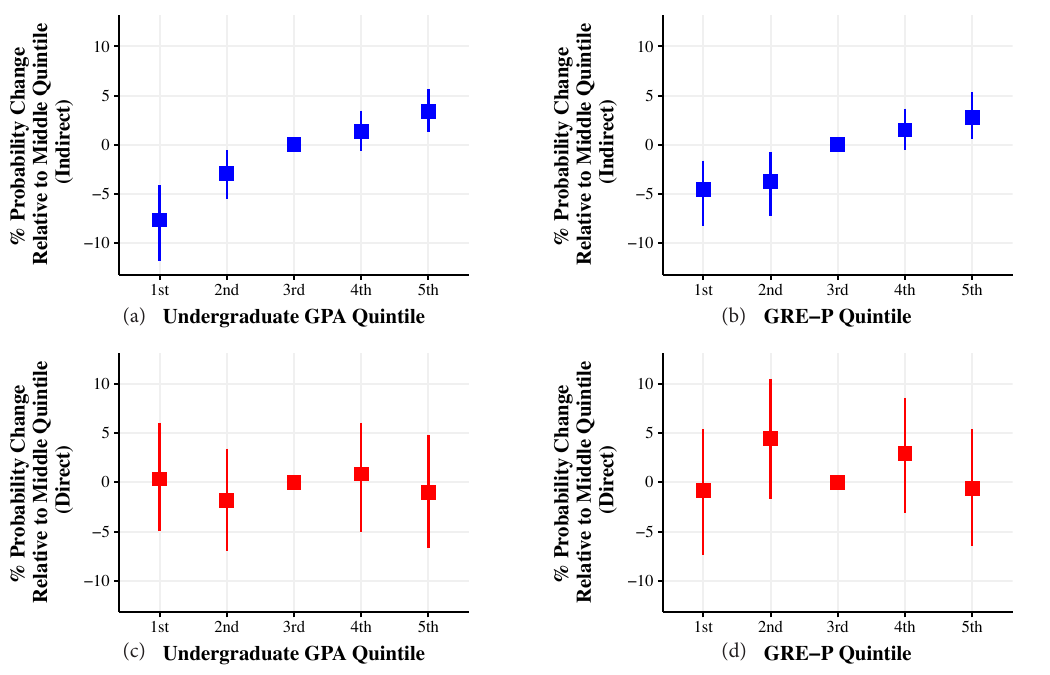}
\caption{\label{fig:mediation_quintile} Plots of the indirect and direct effects for UGPA and GRE-P when treated as categorical predictors.  Results are qualitatively similar to those treating the variables as continuous. (a) Indirect effect of undergraduate GPA when split into quintiles and treated as a categorical variable. (b) Indirect effect of GRE-P when split into quintiles and treated as a categorical variable. (c) Direct effect of undergraduate GPA when split into quintiles and treated as a categorical variable. (d) Direct effect of GRE-P when split into quintiles and treated as a categorical variable.}
\end{figure*}

In addition to the analysis presented in the main text in which we treated UGPA, GRE-P, and GRE-Q as continuous variables, we also performed a mediation analysis in which we turned these predictors into categorical variables by binning scores into quintiles.  This means that students whose UGPA score fell in the top 20\% of our data were classified in the 5th (top) quintile and students whose scores fell in the bottom 20\% were classified as the 1st (bottom) quintile.  We grouped students in this way for their UGPA, GRE-P, and GRE-Q scores separately.  Grouping the variables in this way may provide readers with a more intuitive feeling for how PhD completion probability is predicted to change between students.

Grouping UGPA scores by quintile leads to five possible classifications: 1st quintile ($<3.38$), 2nd quintile (3.38-3.61), 3rd quintile (3.61-3.80), 4th quintile (3.80-3.91), and 5th quintile ($>3.91$).  Similarly for GREP the groupings are: 1st quintile ($<30$), 2nd quintile (30-46), 3rd quintile (46-61), 4th quintile (61-74), 5th quintile ($>74$).  We also explored treating GRE-Q as a categorical variable by splitting students into quintiles, but the scores for physicists taking the GRE-Q are so tightly distributed that we do not believe much intuition is gained from the analysis.  In particular, the second quintile of scores ranges from 89.1-90.9 and the third quintile ranges from 85-89.1, while the fifth quintile includes all scores less than 77.  Since there is such a small score difference between students who scored in bottom and top quintiles on the GRE-Q, we do not believe it is useful to group the scores in this way. 

Results for treating UGPA and GRE-P as categorical variables are qualitatively similar to those presented in the main paper.  Using the ``Medium" group as a reference, the predicted direct and indirect effects associated with UGPA and GRE-P are shown in \ref{fig:mediation_quintile}.  

For UGPA, a student who earns a score in the 5th quintile as opposed to the 1st quintile increases their probability of PhD completion by approximately 10\% due to the indirect effect on GGPA.  We had reported in the main paper that the predicted probability change in completion for a student in the 10th percentile of UGPA scores to the 90th percentile (a change in score from 3.2 to 3.98) was 11\%, which is consistent with the results shown here.  As in the main text, all direct effects of UGPA are not statistically significant. 

Meanwhile the indirect effect of GRE-P on completion predicts a probability change of approximately 6\% for a student who earns a top quintile GRE-P score as opposed to a bottom quintile score.  This is consistent with the main analysis, which estimated that due to the indirect effect of GRE-P on GGPA a change from the 10th  to 90th percentile of GRE-P scores (a change in GRE-P from 21 to 85 among the overall test-taking population) increased the predicted probability of completion by 7\%.  Again, as in the main text, all direct effects of GRE-P are not statistically significant.

\clearpage
\section{Sensitivity Analysis Plots}

\begin{figure*}[h]
    \includegraphics[width=\textwidth]{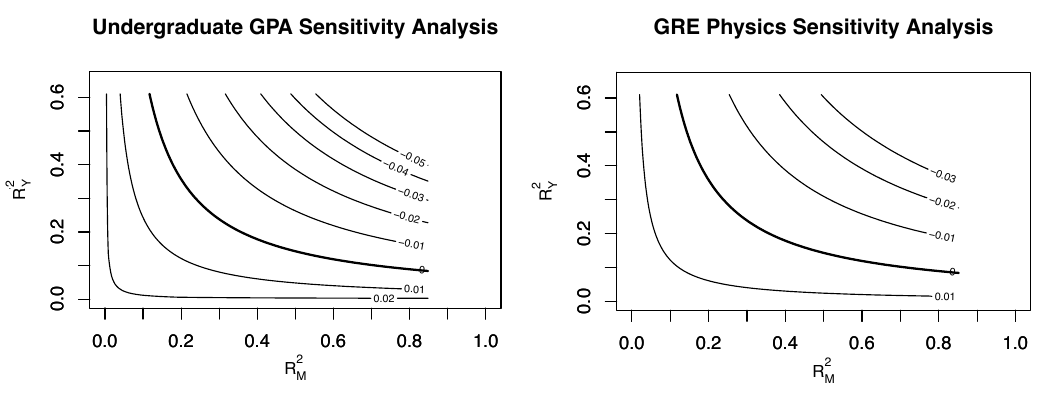}
    \caption{\label{fig:sensitivity} Contour plots of the total variance of the mediator and outcome variables that an unobserved confounder would have to explain in order to make the indirect effects of UGPA and GRE-P vanish.  The bold lines represent solutions to the equations $R^2_M \times R^2_Y = 0.072$ and $R^2_M \times R^2_Y = 0.069$, for UGPA and GRE-P respectively.  These plots are generated using a typical imputed dataset and are not pooled across imputations.  Plots are generated in the \texttt{R} package \texttt{mediation} \cite{mediation_package}.}
\end{figure*}

In the main analysis we summarized the results of sensitivity analyses performed on each of the imputed data sets used in the single mediator models to assess their robustness.  We let $R^2_M$ and $R^2_Y$ represent the proportions of original variances explained by the unobserved confounder for the mediator GGPA and the outcome PhD completion, respectively.  The result of a sensitivity analysis is a single value representing the product $R^2_M \times R^2_Y$ that identifies the amount of original variance in the mediator and outcome that the confounder would have to explain in order to make the observed effect vanish. Hence, the sensitivity analysis results in a family of solutions for which the equation $R^2_M \times R^2_Y = constant$ is satisfied.  

Our sensitivity analysis reveals that for a confounder to explain enough variance to make the indirect effects of UGPA and GRE-P on GGPA vanish, the product $R^2_M \times R^2_Y$ on average would have to be 0.072 and 0.069, respectively.  The standard deviation of the UGPA sensitivity analysis results across all imputed datasets was 0.006.  This indicates there was little difference in the outcome of sensitivity analyses regardless of imputation.  The standard deviation of the GRE-P sensitivity analysis results across all imputed datasets was 0.001; again, there was little difference in sensitivity analysis results across imputations.  

Contour plots of solutions to the equation $R^2_M \times R^2_Y = 0.072$ and $R^2_M \times R^2_Y = 0.069$, for UGPA and GRE-P respectively, are shown in Figure \ref{fig:sensitivity} by the line labeled 0 and in bold.  Other combinations of higher $R^2_M \times R^2_Y$ values are plotted as well, which demonstrate values at which the estimated indirect effect would actually become negative.

\clearpage
\section{Alternative imputation model using outcome variables}
The primary analysis employed the well-known method of multiple imputation \cite{nissen2019missing, van_buuren_flexible_2018} to handle cases of missing UGPA and GRE-P scores.  Multiple imputation handles missing data by replacing the missing values using a statistical model based on the complete data.  This process is performed several times to generate multiple versions of the complete data set in order to reflect uncertainty in the missing data.

The imputation model used in \citeauthor{miller_typical_2019} \cite{miller_typical_2019} predicted students' missing UGPA and GRE-P scores using their GRE-Q, GRE-V, gender, race, and program tier (the other independent variables in the regression model presented in \cite{miller_typical_2019}).  GGPA was not included as a predictor in the imputation model since it was not included as a variable of interest in the analysis.  Thus, including GGPA in the models presented in the current paper presented a choice regarding our approach to handling missing data: to use the same imputation model as \citeauthor{miller_typical_2019} in order to directly compare results across analyses, or to use a slightly different imputation model that also included GGPA as a predictor variable.  Ultimately we opted to keep the same imputation approach across papers, but we present here the results of the paper had we used the alternative imputation model in which GGPA was included.     

It may seem counterintuitive to use GGPA, which was used as an outcome variable in the primary analysis, in a model of data imputation.  Employing a model of data imputation that uses the outcome variable to predict missing values of the independent variable may seem like a self-fulfilling prophecy, guaranteeing a relationship to exist between them.  However, research suggests that including all variables, including the outcome variable, in the imputation model in fact tends to produce less biased results \cite{moons_using_2006, little1992regression, vanGinkel2020misconceptions}.  This further motivates our decision to include this alternative imputation approach, which demonstrates that results are qualitatively similar regardless of which imputation model is used. 

As in the main paper, we begin by exploring the relationships between variables using bivariate correlations.  We then examine the unique predictive effects of different admission metrics on graduate GPA using a multiple linear regression model.  Mediation analysis is used to examine the role that graduate GPA plays in PhD completion by breaking down effects into direct and indirect components.  Lastly, we perform a second set of mediation analyses using multiple mediators in which we compare the relative importance of tier and graduate GPA as mediators in our model of PhD completion.  At each point in the analysis, we provide brief summaries of the important results and point out any differences that arise from the primary analysis due to the use of a different imputation model.

\subsection{\label{subsec:RoleofGGPA_altimp}Exploring the Role of GGPA - Alternative imputation model using outcome variable}

%
%

\begin{table*}[h]
\centering
\resizebox{\textwidth}{!}{%
\def\arraystretch{.75}
\begin{tabular}{lccccccc}
  \hline
 Measure (M + SD) & UGPA & GRE-Q & GRE-V & GRE-P & GGPA & Final Disp. & Gender \\ 
  \hline
  UGPA (3.6 $\pm$ 0.3) & -- & (0.26, 0.35)  & (0.13, 0.23)  & (0.27, 0.37)  & (0.32, 0.41)  & (0.14, 0.23)  & (-0.09, 0.01)  \\ 
  
  GRE-Q (83.3 $\pm$ 10.4) & 0.31 & -- & (0.30, 0.37)  & (0.46, 0.54)  & (0.13, 0.22)  & (0.10, 0.19)  & (-0.16, -0.07)  \\ 
  
  GRE-V (76.3 $\pm$ 18.7) & 0.18 & 0.34 & -- & (0.22, 0.30)  & (0.06, 0.15)  & (0.02, 0.10)  & (-0.02, 0.07)  \\ 
  
  GRE-P (52.9 $\pm$ 23.2) & 0.32 & 0.51 & 0.26 & -- & (0.21, 0.31)  & (0.11, 0.20)  & (-0.33, -0.24)  \\ 
  
  GGPA (3.5 $\pm$ 0.5) & 0.37 & 0.18 & 0.10 & 0.26 & -- & (0.39, 0.46)  & (-0.06, 0.03)  \\ 
  
  Final Disp. & 0.19 & 0.15 & 0.06 & 0.16 & 0.43 & -- & (-0.09, 0.00)  \\
  
  Gender & -0.04 & -0.11 & 0.03 & -0.29 & -0.02 & -0.05 & --  \\ 
  
   \hline
\end{tabular}}
\caption{\label{tab:correlationsALTIMP} A matrix showing bivariate correlations between continuous and dichotomous variables under an alternate imputation model.  Correlations are shown in the lower diagonal while confidence intervals for those correlations are shown in the upper diagonal.  UGPA and GRE-P correlations with GGPA are slightly boosted compared to the correlations presented in the main text.}
\end{table*}

Since only aspects of the analysis including UGPA and GRE-P are influenced by the new imputation model, only correlations involving these two variables are different from the main paper.  For instance, UGPA ($r_{xy}=0.37$, $p<.001$) and GRE-P ($r_{xy}=0.26$, $p<.001$) are again positively correlated with GGPA, meaning that students with higher scores in these metrics tend to earn higher GGPA scores.  However, as expected due to the use of GGPA in this imputation model, the correlation between UGPA and GGPA is now higher ($r_{xy}=0.37$ rather than 0.29).  The correlation between GGPA and GRE-P is also slightly higher ($r_{xy}=0.26$ rather than 0.22).  Thus the boost in correlation is higher for UGPA than GRE-P, indicating that the primary effect of adding GGPA into the imputation model is to boost the effects associated with UGPA. 

Other results reported in Table \ref{tab:correlationsALTIMP} are identical to the main paper.  Again, GGPA is the predictor most strongly correlated with final disposition ($r_{pb}=0.43$).  This value is statistically significant ($p<.001$), meaning that the mean GGPA score of students who do not complete a PhD are statistically different from those who do complete their PhD.  Specifically, the positive correlation means that students with higher GGPA scores are more likely to finish their PhD program successfully.     

\begin{table}[H]
\centering
\def\arraystretch{.75}
\begin{tabular}{llcc}
\multicolumn{4}{c}{Multiple Regression Results (*$p<.05$, **$p<.01$)}                                                                       \\ \hline\hline
\multicolumn{2}{l}{}                            & \multicolumn{2}{c}{Dependent Variable - GGPA} \\ \cline{3-4} 
\multicolumn{2}{l}{\begin{tabular}[c]{@{}l@{}}Independent\\ Variable\end{tabular}} &
  \begin{tabular}[c]{@{}c@{}}Coefficient\\ (Standard Error)\end{tabular} &
  \begin{tabular}[c]{@{}c@{}}Standardized\\ Coefficient\end{tabular} \\ \hline

\rowcolor[gray]{.9}[\tabcolsep]   
\multicolumn{2}{l}{Intercept} & 1.51** (0.15) & -0.08** \\

\rowcolor[gray]{.9}[\tabcolsep] 
\multicolumn{2}{l}{UGPA} & 0.50** (0.04) & 0.34** \\

\rowcolor[gray]{.9}[\tabcolsep] 
\multicolumn{2}{l}{GRE-P} & $40 \times 10^{-4}$** ($6 \times 10^{-4}$) & 0.19**          \\

\rowcolor[gray]{.9}[\tabcolsep] 
\multicolumn{2}{l}{GRE-Q} & $-3 \times 10^{-4}$ ($12 \times 10^{-4}$) & -0.01              \\

\rowcolor[gray]{.9}[\tabcolsep] 
\multicolumn{2}{l}{GRE-V} & $-2 \times 10^{-6}$ ($6 \times 10^{-4}$) & 0.00              \\

\multicolumn{2}{l}{Black} & -0.10 (0.09) & -0.20              \\

\multicolumn{2}{l}{Hispanic} & 0.01 (0.07) & 0.01  \\

\multicolumn{2}{l}{Native Am.} & 0.07 (0.24) & 0.16  \\

\multicolumn{2}{l}{Asian}  & -0.03 (0.07)  & -0.05  \\

\multicolumn{2}{l}{Other} & -0.24* (0.10)  & -0.50*  \\

\multicolumn{2}{l}{Undisclosed} & 0.10** (0.02)  & 0.20**  \\

\multicolumn{2}{l}{Gender} & 0.07* (0.03)  & 0.15* \\ \hline\hline

\multicolumn{2}{l}{\textit{N}} & 1955 &          \\

\multicolumn{2}{l}{Adjusted \textit{R}-Squared} & 0.17                      &             
\end{tabular}
\caption{\label{tab:regressionALTIMP} Results of a multiple regression for alternative imputation model.  As in the correlation matrix, the coefficients of the UGPA and GRE-P terms are boosted compared to the main analysis.  This effect is stronger for UGPA.  Also notable is that the GRE-Q coefficient is relatively weaker compared to the main analysis.}
\end{table}

The results of the multiple regression analysis predicting GGPA shown in Table \ref{tab:regressionALTIMP} are qualitatively consistent with those presented in the main analysis.  Significant unique predictive effects at the 95\% threshold were found for the numerical metrics UGPA and GRE-P, meaning that students with higher UGPA and GRE-P scores tended to received higher GGPA scores.  GRE-Q does not reach the threshold for statistical significance.

The alternative imputation model increased the effects associated with UGPA ($\beta = 0.34$, $t=12.67$, $p<.01$) and GRE-P ($\beta = 0.19$, $t=6.18$, $p<.01$).  Specifically, the UGPA standardized coefficient increased from 0.24 to 0.34 and the GRE-P standardized coefficient increased from 0.15 to 0.19.  As with the bivariate correlations, the alternate imputation model boosts the effect of UGPA on GGPA more than it does for GRE-P.  Furthermore, the overall model fit improved under the alternate imputation approach ($R^2$ increased from 0.11 to 0.17).

\clearpage
\subsection{\label{subsec:MediationResults_altimp}Mediation Analysis - Alternative imputation model using outcome variable}

\begin{figure*}[h]
    \includegraphics[width=\textwidth]{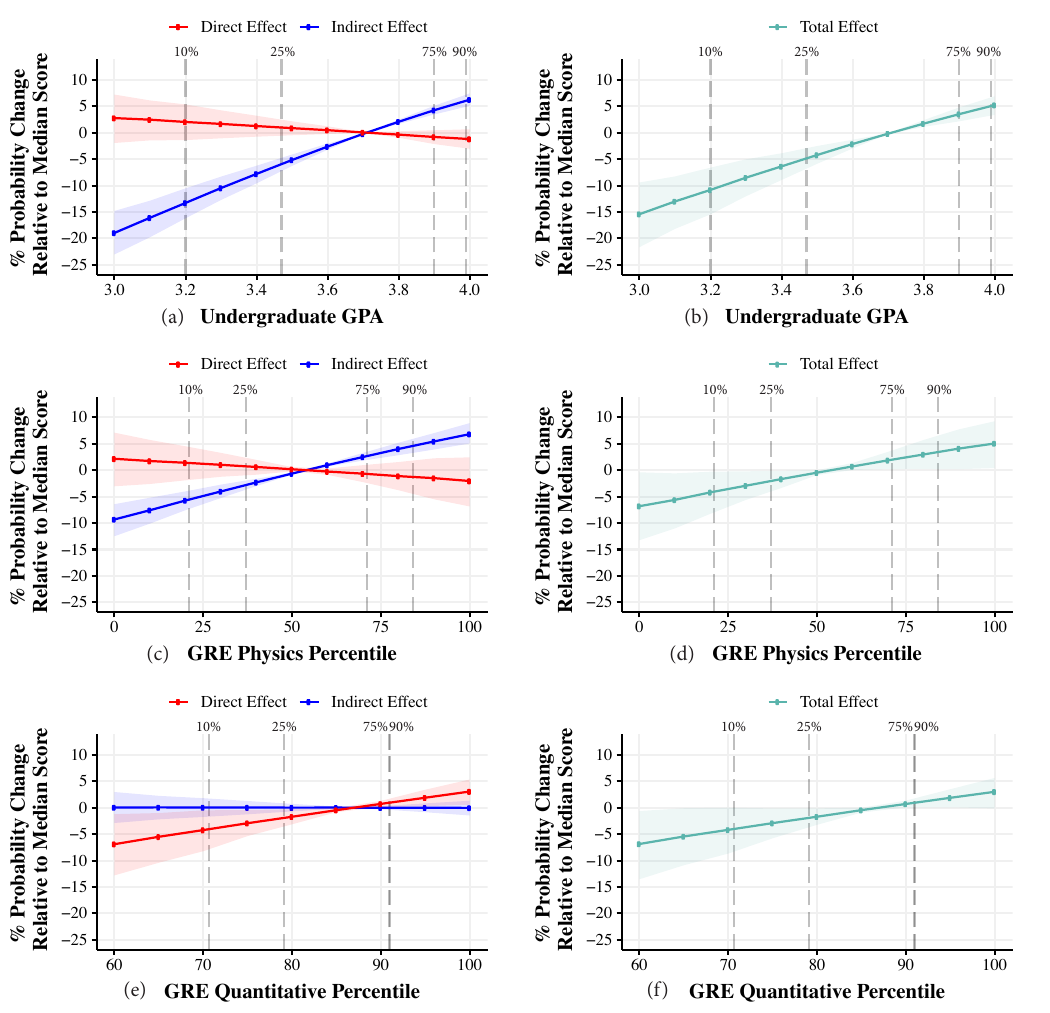}
    \caption{\label{fig:mediationsALTIMP} Mediation analysis results from an alternate imputation model predicting a student's change in probability of completion, split into direct (red) and indirect via graduate GPA (blue) effects. The indirect effects of UGPA and GRE-P are larger compared with the main analysis, and again UGPA is clearly a stronger predictor in these models.  Also notable is that the total effect of GRE-P just meets the threshold of significance under this model.  The direct effect of GRE-Q shifts to significant at 95\% threshold as well.}
\end{figure*}

Figure \ref{fig:mediationsALTIMP} graphically summarizes the results of the mediation analysis under the alternate imputation model.  As in the main paper, Figures \ref{fig:mediationsALTIMP}a), c), and e) display the results of the three separate analyses predicting PhD completion probability changes relative to the median value of each independent variable.  Points on each plot represent individual calculations of probability change relative to each admission metric's median value.  Hence, the median value of the $x$-axis variable is clearly shown on the plot as the point where the direct (red) and indirect effect (blue) lines intersect, corresponding to a total probability change of 0\%.  Percent probability changes due to direct effects of each admissions metric on PhD completion are shown in red, while indirect effects on PhD completion transmitted through GGPA are shown in blue.  Plots of the total effect of each variable on PhD completion probability, which is the sum of the direct and indirect effects, are also shown in Figures \ref{fig:mediationsALTIMP}b), d), and f). The shaded ribbons around the lines representing the best-estimates of probability change show the 95\% confidence interval.  Hence, if this shaded region contains the line $y = 0$,  the effect is not statistically significant at the $\alpha=.05$ level.  

For the most part the results presented in Figure \ref{fig:mediationsALTIMP} are qualitatively consistent with those in the main analysis.  The primary message from this model and the main text is that among common quantitative admissions metrics, undergraduate GPA offers the most promising insight into whether physics graduate students will earn a PhD.    

Under the alternate imputation model, there is a noticeable increase in both the total and indirect predictive effects associated with UGPA.  

While the total effect of GRE-P just barely missed reaching the threshold for statistical significance in the main paper, under the alternate imputation model the shaded ribbons around total effect of GRE-P  no longer cross the line $y=0$. Hence, the total effect of GRE-P meets the threshold of statistical significance at the $\alpha=0.05$ level in this model, although it is again very close.  The magnitude of the total effect is similar to the main paper (a change from the 10th  to 90th percentile of GRE-P scores among students in this model increases the predicted probability of completion by approximately 8\% in this model as opposed to 7\% in the main paper).  This is likely due to the slight boost in the magnitude of the GRE-P's indirect effect on completion; a change from the 10th  to 90th percentile of GRE-P scores among students in this model increases the predicted probability of completion by approximately 10\%, up from 8\% in the main analysis.

In contrast to the main analysis, the direct effect of GRE-Q under the alternate imputation model meets the threshold of statistical significance at the $\alpha=0.05$ level; the indirect predictive effects of GRE-Q on PhD completion remain statistically insignificant.  

Their sum, the total effect, also barely meets the threshold for significance.  This is consistent with \cite{miller_typical_2019}, in which the predictive effect of GRE-Q on completion was significant but was not strong and in fact bordered on insignificance as well ($p=0.048$). 

Given the weak relationship between GRE-Q and GGPA revealed by the multiple regression analysis in Section \ref{subsec:RoleofGGPA_altimp}, it is expected that the indirect effect shown in blue on Figure \ref{fig:mediationsALTIMP} is nearly zero.  Indeed, any predictive effect from GRE-Q on completion appears to stem from the direct effect of GRE-Q on PhD completion.


\clearpage
\subsection{\label{subsec:tierMediation_altimp} Multiple Mediators - Alternative imputation model using outcome variable}

\begin{figure*}[h]
    \includegraphics[width=\textwidth]{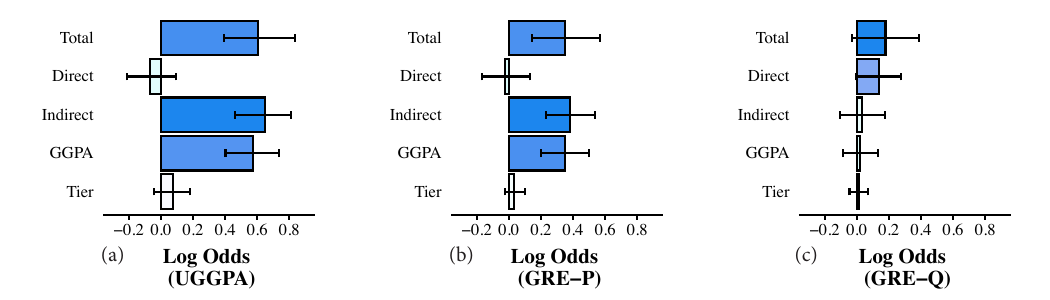}
    \caption{\label{fig:joint_mediationsALTIMP} Multiple mediator results under an alternate imputation model.  Results are qualitatively consistent with the single mediator case.  As in the main analysis, the indirect effects transmitted through program tier are not statistically significant in any of the multiple mediator models and are minuscule compared to the effects associated with GGPA.}
\end{figure*}

As discussed in the main text, methods to disentangle the effect that individual paths in a multiple mediator model have on the dependent variable has been the subject of much recent research \cite{daniel_causal_2015, imai_identification_2013}.  For the alternate imputation model we again follow the method provided in \cite{steen_flexible_2017}, which extends the counterfactual mediation analysis framework to deal with particular cases of multiple mediators.  The results of this analysis are direct and indirect effects of each admissions metric on PhD completion, with the indirect effect broken down into the portion transmitted through GGPA and the portion transmitted through program tier.  

The output is again qualitatively similar to the single mediator case described previously: direct and indirect effects are calculated, and effects for which the 95\% error bars do not cross zero are statistically significant.  These results are summarized in Figure \ref{fig:joint_mediationsALTIMP}.  The single mediator results presented in Section \ref{subsec:MediationResults_altimp} are fully consistent qualitatively with the results of this multiple mediator analysis. Again, both UGPA and GRE-P have a significant total effect on PhD completion, and that effect is fully transmitted through GGPA.  However the effect of UGPA is larger than the effect of GRE-P.  The increase in log-odds associated with a one standard deviation increase in UGPA is 0.60 and GRE-P is 0.35.  As in the single mediator case the direct effect of GRE-Q is just barely significant.  

Despite its inclusion, the indirect effects transmitted through program tier are not statistically significant in any of the multiple mediator models and are minuscule compared to the effects associated with GGPA.

\clearpage
\section{Non-US Results}

In order to focus on issues of diversity and inclusion associated most strongly with US applicants, the main paper only used data from the subset of domestic graduate students.  This decision was also motivated by the the markedly different distributions of scores for US and Non-US students.  Here, we repeat the analyses presented in the main paper on the subset of international graduate students.

To match the primary analysis, we first present the relationships between variables using bivariate correlations.  We then examine the unique predictive effects of different admission metrics on international students' graduate GPA using a multiple linear regression model.  A mediation analysis is used to examine the role that graduate GPA plays in PhD completion by international students by breaking down effects into direct and indirect components.  Lastly, we perform a second set of mediation analyses using multiple mediators in which we compare the relative importance of tier and graduate GPA as mediators in our model of PhD completion by international students.  At each point in the analysis, we provide brief summaries of the important results and point out any differences between results for US and Non-US students.

\subsection{\label{subsec:RoleofGGPA_nonus}Exploring the Role of GGPA - NonUS}

%
%

\begin{table*}[h]
\centering
\resizebox{\textwidth}{!}{%
\def\arraystretch{.75}
\begin{tabular}{lccccccc}
  \hline
 Measure (M + SD) & UGPA & GRE-Q & GRE-V & GRE-P & GGPA & Final Disp. & Gender \\ 
  \hline
  UGPA (3.6 $\pm$ 0.4) & -- & (0.05, 0.18)  & (0.07, 0.20)  & (0.17, 0.34)  & (0.05, 0.17)  & (-0.03, 0.09)  & (-0.09, 0.02)  \\ 
  
  GRE-Q (87.2 $\pm$ 7.4) & 0.12 & -- & (0.25, 0.34)  & (0.47, 0.56)  & (0.09, 0.19)  & (0.00, 0.11)  & (-0.21, -0.11)  \\ 
  
  GRE-V (60.5 $\pm$ 29.1) & 0.14 & 0.29 & -- & (0.27, 0.37)  & (0.07, 0.18)  & (-0.04, 0.06)  & (-0.29, -0.16)  \\ 
  
  GRE-P (75.1 $\pm$ 22.3) & 0.26 & 0.52 & 0.32 & -- & (0.15, 0.27)  & (0.00, 0.11)  & (-0.29, -0.16)  \\ 
  
  GGPA (3.7 $\pm$ 0.4) & 0.11 & 0.14 & 0.13 & 0.21 & -- & (0.21, 0.31)  & (-0.05, 0.05)  \\ 
  
  Final Disp. & 0.03 & 0.05 & 0.01 & 0.05 & 0.26 & -- & (-0.08, 0.03)  \\
  
  Gender & -0.04 & -0.16 & -0.06 & -0.23 & 0.00 & -0.02 & --  \\ 
  
   \hline
\end{tabular}}
\caption{\label{tab:correlationsNONUS} A matrix showing bivariate correlations between continuous and dichotomous variables for international students.  Correlations are shown in the lower diagonal while confidence intervals for those correlations are shown in the upper diagonal.  GGPA persists as the best predictor of PhD completion, but none of the other quantitative metrics (UGPA, GRE-P, GRE-Q, and GRE-V) have a statistically significant correlation with PhD completion.}
\end{table*}

\begin{figure}
\includegraphics[]{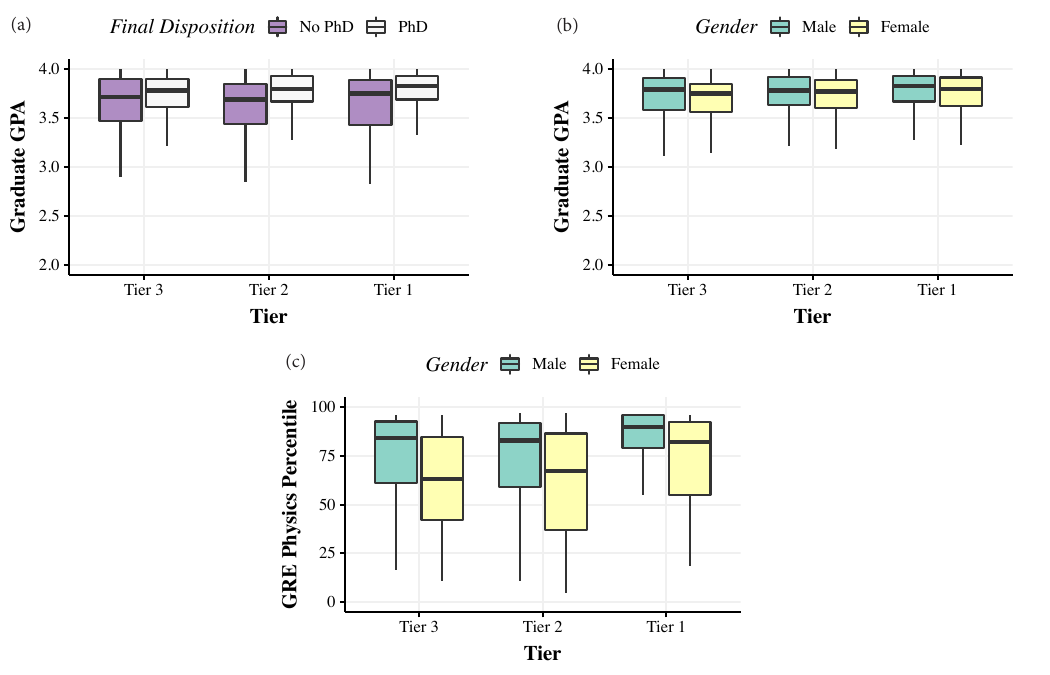}
\caption{\label{fig:boxplotsNONUS} (a) Graduate GPA by program tier and final disposition. International students who do not complete a PhD earn lower graduate grades than students who complete their programs. (b) Graduate GPA by program tier and gender. Male and female international students earn similar graduate grades, and this trend holds across program tier. (c) GRE Physics by program tier and gender. As for US students, across all program tiers there is a significant gap in scores between male and female GRE-P test takers.}
\end{figure}

As was the case for US students, GGPA is the predictor most strongly correlated with final disposition ($r_{pb}=0.26$) for international students.  This value is statistically significant ($p<.001$), meaning that the mean GGPA score of students who do not complete a PhD are statistically different from those who do complete their PhD.  Specifically, the positive correlation means that students with higher GGPA scores are more likely to finish their PhD program successfully.  This trend is visually apparent in Figure \ref{fig:boxplotsNONUS}(a), which shows boxplots of GGPA grouped by PhD completers and non-completers.

There is no statistical difference between average GGPA scores for male and female international students ($r_{pb}=0.00$, $p=0.98$), as demonstrated in Figure \ref{fig:boxplotsNONUS}(b).  However, the statistically significant difference between males and females in GRE-P performance persists ($r_{pb}=-0.23$, $p<.001$).  Thus the gap in GRE-P performance by gender appears in both the US and NonUS groups, but disappears in subsequent graduate performance metrics. 

In contrast to US students, none of the quantitative metrics (UGPA, GRE-P, GRE-Q, and GRE-V) have a statistically significant correlation with PhD completion.  This already suggests that the results of subsequent mediation analyses are likely to differ between the US and NonUS student groups.  Also different are the results of one-way independent ANOVA tests, which show that the main effect of program tier on GGPA is insignificant, $F(2, 1448) = 2.80$, $p=0.06$. This indicates that GGPA scores of NonUS students do not differ regardless of the tier of their institution, whereas GGPA scores of US students tended to be higher as program tier increased.

\begin{table}
\centering
\def\arraystretch{.75}
\begin{tabular}{llcc}
\multicolumn{4}{c}{Multiple Regression Results (*$p<.05$, **$p<.01$)}                                                                       \\ \hline\hline
\multicolumn{2}{l}{}                            & \multicolumn{2}{c}{Dependent Variable - GGPA} \\ \cline{3-4} 
\multicolumn{2}{l}{\begin{tabular}[c]{@{}l@{}}Independent\\ Variable\end{tabular}} &
  \begin{tabular}[c]{@{}c@{}}Coefficient\\ (Standard Error)\end{tabular} &
  \begin{tabular}[c]{@{}c@{}}Standardized\\ Coefficient\end{tabular} \\ \hline

\rowcolor[gray]{.9}[\tabcolsep]   
\multicolumn{2}{l}{Intercept} & 3.03** (0.17) & -0.02 \\

\rowcolor[gray]{.9}[\tabcolsep] 
\multicolumn{2}{l}{UGPA} & 0.06 (0.03) & 0.06 \\

\rowcolor[gray]{.9}[\tabcolsep] 
\multicolumn{2}{l}{GRE-P} & $29 \times 10^{-4}$** ($7 \times 10^{-4}$) & 0.17**          \\

\rowcolor[gray]{.9}[\tabcolsep] 
\multicolumn{2}{l}{GRE-Q} & $20 \times 10^{-4}$ ($16 \times 10^{-4}$) & 0.04              \\

\rowcolor[gray]{.9}[\tabcolsep] 
\multicolumn{2}{l}{GRE-V} & $7 \times 10^{-4}$ ($4 \times 10^{-4}$) & 0.05              \\

\multicolumn{2}{l}{Gender} & 0.05 (0.03)  & 0.12 \\ \hline\hline

\multicolumn{2}{l}{\textit{N}} & 1451 &          \\

\multicolumn{2}{l}{Adjusted \textit{R}-Squared} & 0.05                      &             
\end{tabular}
\caption{\label{tab:regressionNONUS} Coefficients of a multiple regression analysis modeling graduate GPA as a function of common quantitative admissions metrics for international student data.  Reference category for gender is male.}
\end{table}

The results obtained in our multiple regression analysis predicting GGPA are summarized in Table \ref{tab:regressionNONUS}.  Significant unique predictive effects at the 95\% threshold were found only for the numerical metric GRE-P ($\beta = 0.17$, $t=4.42$, $p<.01$).  Hence, international students with higher GRE-P scores tended to received higher GGPA scores.  As in the main text, the GRE-Q coefficient is not statistically significant ($\beta = 0.04$, $t=1.20$, $p=0.23$) and is less than a quarter the magnitude of the GRE-P coefficient.  Combined with the low correlation between GRE-Q and PhD completion, this result indicates that GRE-Q scores likely will not have significant predictive effects in subsequent mediation analyses.  

In contrast with the US student group, the UGPA coefficient is statistically insignificant ($\beta = 0.06$, $t=1.69$, $p=0.09$) and is approximately 33\% the size of the GRE-P coefficient.  This suggests UGPA has comparatively little predictive effect on graduate course performance for international students.  Similar to the GRE-Q, this result suggests that UGPA will not not have significant predictive effects in subsequent mediation analyses.

\begin{table}
\centering
\def\arraystretch{.75}
\begin{tabular}{llcc}
\multicolumn{4}{c}{Logistic Regression Results (*$p<.05$, **$p<.01$)}                                                                       \\ \hline\hline
\multicolumn{2}{l}{}                            & \multicolumn{2}{c}{Dependent Variable - Final Disposition} \\ \cline{3-4} 
\multicolumn{2}{l}{\begin{tabular}[c]{@{}l@{}}Independent\\ Variable\end{tabular}} &
  \begin{tabular}[c]{@{}c@{}}Coefficient\\ (Standard Error)\end{tabular} &
  \begin{tabular}[c]{@{}c@{}}Standardized\\ Coefficient\end{tabular} \\ \hline

\rowcolor[gray]{.9}[\tabcolsep]   
\multicolumn{2}{l}{Intercept} & -0.37 (0.96) & 1.19** \\

\rowcolor[gray]{.9}[\tabcolsep] 
\multicolumn{2}{l}{UGPA} & 0.12 (0.19) & 0.04 \\

\rowcolor[gray]{.9}[\tabcolsep] 
\multicolumn{2}{l}{GRE-P} & $32 \times 10^{-4}$ ($37 \times 10^{-4}$) & 0.07          \\

\rowcolor[gray]{.9}[\tabcolsep] 
\multicolumn{2}{l}{GRE-Q} & 0.01 ($94 \times 10^{-4}$) & 0.08              \\

\rowcolor[gray]{.9}[\tabcolsep] 
\multicolumn{2}{l}{GRE-V} & $-13 \times 10^{-4}$ ($23 \times 10^{-4}$) & -0.04              \\

\multicolumn{2}{l}{Gender} & -0.07 (0.16)  & -0.07 \\ \hline\hline

\multicolumn{2}{l}{\textit{N}} & 1451 &                    
\end{tabular}
\caption{\label{tab:logisticregressionNONUS} Coefficients of a logistic regression analysis modeling PhD completion as a function of common quantitative admissions metrics for international student data.  Reference category for gender is male. For comparison, 25\% of US students (494 of 1955) in this data set did not earn a PhD while 24\% of international students (344 of 1451) in this data set did not earn a PhD.}
\end{table}

Since \cite{miller_typical_2019} did not include the results of a logistic regression model predicting PhD outcome for international students, we include the results of that analysis here in Table \ref{tab:logisticregressionNONUS}.  We observe that none of the quantitative admissions metrics meet the threshold of statistical significance at the $\alpha = 0.05$ level, although GRE-P ($\beta = 0.07$, $p=0.49$) and GRE-Q ($\beta = 0.08$, $p=0.19$) are closest.

\subsection{\label{subsec:MediationResults_nonus}Mediation Analysis - NonUS}

\begin{figure*}[h]
    \includegraphics[width=\textwidth]{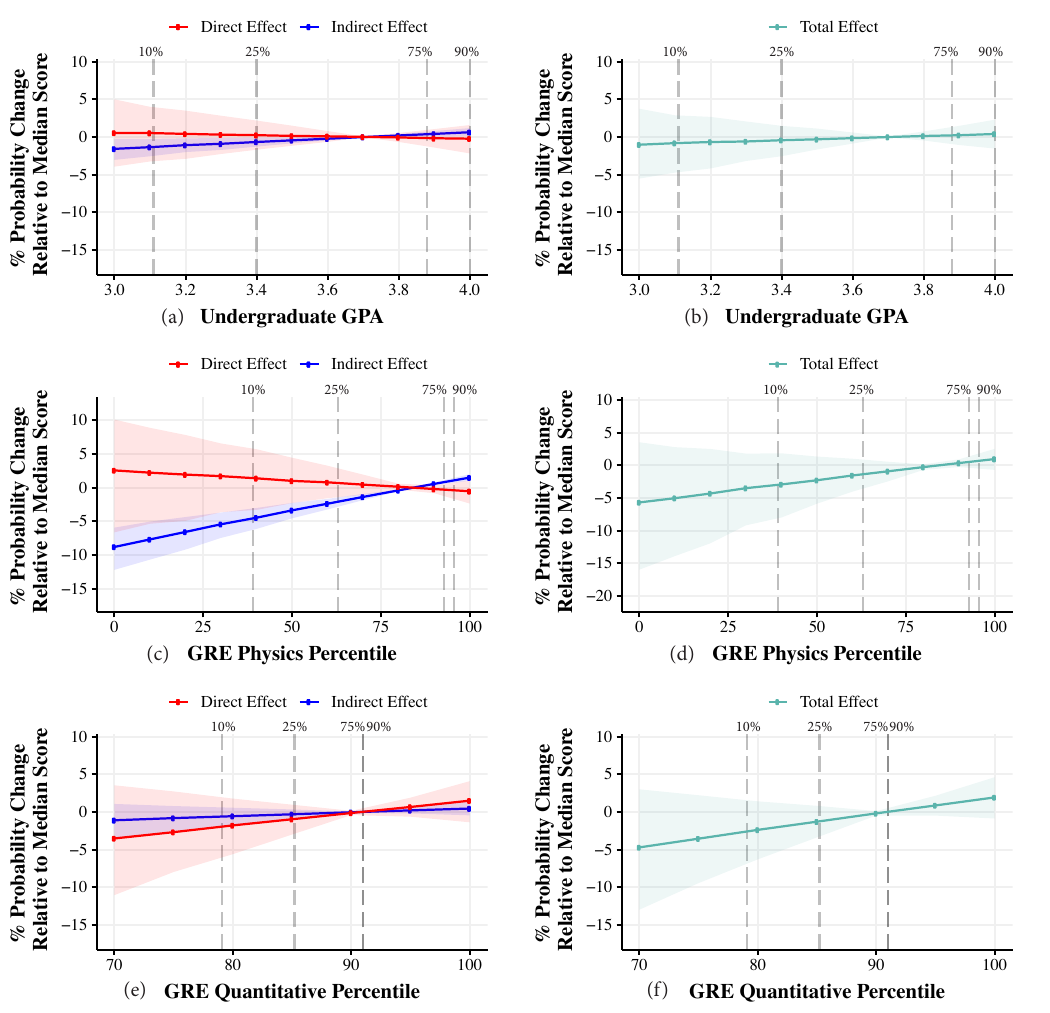}
    \caption{\label{fig:mediationsNONUS} Mediation analysis results using international student data predicting a student's change in probability of completion, split into direct (red) and indirect via graduate GPA (blue) effects.  Only the GRE-P mediation analysis results for international students is similar to that of the US student group.}
\end{figure*}

Figure \ref{fig:mediationsNONUS} graphically summarizes the results of the mediation analysis for the subset of international students.  As in previous discussions, Figures \ref{fig:mediationsNONUS}a), c), and e) display the results of the three separate analyses predicting PhD completion probability changes relative to the median value of each independent variable.  Points on each plot represent individual calculations of probability change relative to each admission metric's median value.  Hence, the median value of the $x$-axis variable is clearly shown on the plot as the point where the direct (red) and indirect effect (blue) lines intersect, corresponding to a total probability change of 0\%.  Percent probability changes due to direct effects of each admissions metric on PhD completion are shown in red, while indirect effects on PhD completion transmitted through GGPA are shown in blue.  Plots of the total effect of each variable on PhD completion probability, which is the sum of the direct and indirect effects, are also shown in Figures \ref{fig:mediationsNONUS}b), d), and f). The shaded ribbons around the lines representing the best-estimates of probability change show the 95\% confidence interval.  Hence, if this shaded region contains the line $y = 0$,  the effect is not statistically significant at the $\alpha=.05$ level.  

Several of the results presented in Figure \ref{fig:mediationsNONUS} for international students are markedly different from the US student population.  

Whereas for US students the total effect of UGPA on PhD completion was statistically significant and almost entirely attributable to the indirect effect of UGPA through GGPA, the total effect of UGPA on completion is not significant for international students.  The estimated probability change in completion for a student in the 10th percentile of UGPA scores to the 90th percentile (a change in score from 3.1 to 3.99) is just 3\%, compared with 11\% for US students.

The confidence ribbon around the estimate of the total effect of GRE-Q for NonUS students (Figure \ref{fig:mediationsNONUS}f) is larger than that for US students, indicating a larger degree of uncertainty around this estimate.  While the total effect of GRE-Q bordered on statistical significance for the US student subset, the wider confidence interval for the NonUS student estimate easily includes the line $y=0$ and therefore does not provide strong evidence that GRE-Q effectively predicts NonUS student PhD completion.  The direct and indirect effect estimates of GRE-Q on completion do not meet the threshold of statistical significance either.

The GRE-P mediation analysis results for international students is similar to that of the US student group.  Again, the total effect of GRE-P barely misses the threshold for statistical significance (Figure \ref{fig:mediationsNONUS}d).  However, any effect of GRE-P on a student's PhD completion appears to be entirely mediated by GGPA, defined by the fact that the indirect effect of GRE-P is statistically significant across all magnitudes of score change while its direct effects are not.  The model estimates that due to the indirect effect of GRE-P on GGPA, a change from the 10th  to 90th percentile of GRE-P scores among international students in this study (a change in GRE-P from 21 to 85 among the overall test-taking population) the predicted probability of completion increases by 8\%.


\clearpage
\subsection{\label{subsec:tierMediation_nonus} Multiple Mediators - NonUS}

\begin{figure*}[h]
    \includegraphics[width=\textwidth]{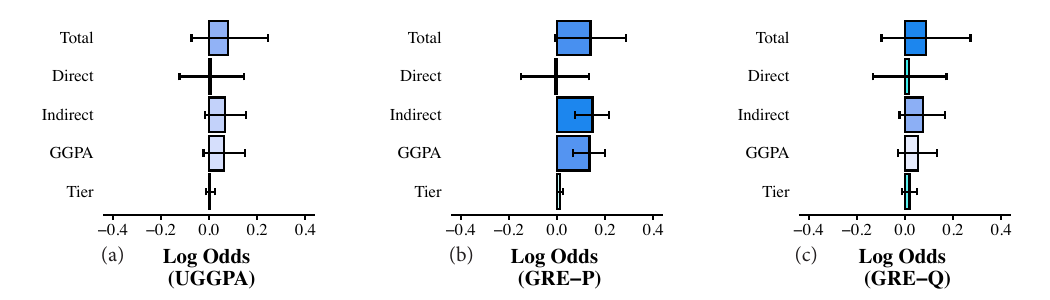}
    \caption{\label{fig:joint_mediationsNONUS} Multiple mediator results for NonUS students.  As in the main analysis, the indirect effects transmitted through program tier are not statistically significant in any of the multiple mediator models and are minuscule compared to the effects associated with GGPA.  Results are mostly consistent with the single mediator case, although in this GRE-Q model the total indirect effect is slightly larger than the direct effect.  However the large error bars indicate these estimates are highly uncertain.}
\end{figure*}

Methods to disentangle the effect that individual paths in a multiple mediator model have on the dependent variable has been the subject of much recent research \cite{daniel_causal_2015, imai_identification_2013}.  For the international student model we again follow the method provided in \cite{steen_flexible_2017}, which extends the counterfactual mediation analysis framework to deal with particular cases of multiple mediators.

Again despite its inclusion, the indirect effects transmitted through program tier are not statistically significant in any of the multiple mediator models and are minuscule compared to the effects associated with GGPA.

Direct and indirect effects are calculated for the multiple mediator mode; effects for which the 95\% error bars do not cross zero are statistically significant.  These results are summarized in Figure \ref{fig:joint_mediationsNONUS}.  The single mediator results presented in Section \ref{subsec:MediationResults_nonus} are mostly consistent qualitatively with the results of this multiple mediator analysis. None of the metrics have a statistically significant total effect, although GRE-P is close.

The primary difference between the single mediator and multiple mediator models is that in the multiple mediator GRE-Q model, the total indirect effect is slightly larger than the direct effect, whereas this is reversed in the single mediator case.  Still, neither effect is statistically significant in either the single or multiple mediator models.  The confidence intervals for the direct effect are large and clearly cross zero in both models, so this discrepancy is not inconceivable.

\bibliography{supplement.bib}